\documentclass[preprint,12pt]{elsarticle}

\usepackage{amssymb}
\usepackage{amsmath}
\usepackage{graphicx}
\usepackage{subfigure}
\usepackage{array}
\usepackage{xcolor}
\usepackage{soul}
\usepackage{setspace}
\usepackage{ulem}
\usepackage{mathtools}
\usepackage[ruled]{algorithm2e}

\definecolor{dgreen}{rgb}{0,0.6,0}
\begin{document}

\begin{frontmatter}

\title{A Multigroup Moment-Accelerated Deterministic Particle Solver for Time-dependent Thermal Radiative Transfer Problems}

\author[T3]{H. ~Park\corref{cor1}}
\ead{hkpark@lanl.gov}
\cortext[cor1]{Corresponding author Tel.:+1 (505)665-0633}
\address[T3]{
Fluid Dynamics and Solid Mechanics Group(T-3)\\
Los Alamos National Laboratory\\
Los Alamos, NM 87545\\
}

\author[T5]{L.~Chac\'{o}n}
\ead{chacon@lanl.gov}
\address[T5]{
Applied Mathematics and Plasma Physics group (T-5)\\
Los Alamos National Laboratory\\
Los Alamos, NM 87545\\
}

\author[T3]{A. ~Matsekh}
\ead{matsekh@lanl.gov}

\author[T5]{G. ~Chen}
\ead{gchen@lanl.gov}

\begin{abstract}
We propose an efficient, robust, Lagrangian (characteristic-based) transport solver for the time-dependent thermal radiative Transfer (TRT) applications within the context of a moment-accelerated  (High-Order/Low-Order, HOLO) algorithm. This novel transport algorithm inherits the best features of both particle methods (e.g., time accuracy, phase-space adaptivity, positivity) and deterministic, grid-based methods (e.g., no stochastic noise). Particles are evolved by the method of characteristics, with a time-dependent weight that accounts for stiff absorption and reemission process self-consistently. As a result, this approach is able to obtain accurate results while employing large time steps and a moderate number of particles (compared to Implicit Monte Carlo). 
\end{abstract}


\end{frontmatter}
%
%
\section{Introduction/motivation}
\label{sec:intro}
Modeling of high-energy density physics phenomena has important applications in astrophysics and inertial
confinement fusion \cite{castor2004radiation}.
In these applications, thermal X-ray radiation becomes the dominant energy transfer mechanism. Radiative transfer is difficult to model due to the stiff nonlinear interactions between radiation and the host material, which can be described by absorption-reemission physics.
Thermal radiative transfer (TRT) without physical scattering can be described by the following equations,
\begin{align}
\frac{1}{c} \frac{\partial I}{\partial t} + \boldsymbol{\Omega} \cdot \nabla I + \sigma I &= \sigma
B,\label{eq:trt}\\
\rho c_v\frac{\partial T}{\partial t} + \sigma acT^4 - \sigma cE &= 0, \label{eq:tmat}
\end{align}
where $I(\mathbf{r},\boldsymbol{\Omega}, \nu,t)$ is the specific intensity of radiation at location
$\mathbf{r}$, traveling along direction $\boldsymbol{\Omega}$, with speed $c$ and frequency $\nu$ at time $t$, and $E(\mathbf{r},t)=\frac{1}{c} \int_{4\pi}\int_0^\infty I d\nu d\Omega$ is the (gray) radiation energy density.
Here, $\sigma,~\rho,$ and $c_v$ are the opacity, density and specific heat of host media, and  $B(\nu,T)$ is
the Planck function for black body radiation at frequency $\nu$ and material temperature $T$,
\begin{equation}
B(\nu,T) \equiv \frac{2 h \nu^3}{c^2}\frac{1}{e^{h\nu/kT}-1}. \label{eq:planck}
\end{equation}
The TRT system is difficult to solve numerically due to
its large dimensionality and stiff nonlinear coupling between radiation and material through absorption and reemission. Over the last several decades, the development of efficient numerical
algorithms has been an active area of research. 

A popular mitigation strategy for the stiff nonlinear coupling is linearization of
the absorption-reemission physics, such as the Fleck-Cummings linearization in the Implicit Monte Carlo (IMC) method \cite{fl71}.
Another popular discretization is the discrete ordinate ($S_N$) method. The advantage of $S_N$ over IMC is that it is able
to obtain the asymptotic diffusion limit \cite{larsen:1974,larsen:1987} in a relatively straightforward manner \cite{ad01,ad97}.

Another class of discretization techniques, which has gained popularity in recent years, is the method of characteristics (MOC)
\cite{fe16,askew1972characteristics,Hoffman2016696}. MOC solves the transport equation analytically along the particle trajectory (characteristics).  MOC is widely used for solving the neutron transport equation, and it is especially effective when accelerated by the coarse-mesh finite-difference (CMFD) method \cite{sm02}. A similar concept is employed in the MULTI2D radiation-hydrodynamics code \cite{Ramis2009977}. However, the MULTI2D utilizes the "short" characteristic method,
which averages the angular intensity at each surface of the mesh.
Recently, an asymptotic preserving, short characteristic method was proposed  \cite{Sun2015265,Sun2015222}. The so-called asymptotic preserving unified gas kinetics scheme (AP-UGKS) utilizes nonlinear elimination and the integral form of the transport equation, enabling a time-implicit discretization of the collisional source term without operator splitting between advection and collision. However, because their formulation assumes the cell-face averaged radiative flux being the function of the neighboring cells, the method is limited to using an explicit CFL time-step size. 
A space-time characteristics method for solving the neutral-particle transport equations has been proposed in Ref. \cite{pandya:2009}. In their work, the temporal variable is treated as another dimension in order to perform ``$n+1$'' dimensional ray-tracing.

The method we propose in this paper is called "deterministic particle"(DP) method. The term ``deterministic particle'' has been introduced in several previous studies \cite{Degond1990,Pareschi2001,Dimarco2011, Dimarco2013680}. In these DP methods, the transport equation is solved via operator-splitting. The first step treats the advection (streaming) term analytically via the particle equation of the motion, and subsequently the collision physics are solved. Thus, it introduces operator-splitting error.

Our approach is similar to AP-UGKS and the space-time characteristics method in that it avoids operator splitting and analytically integrates temporal evolution of particle weights.  However, our approach combines the DP method with a High-Order, Low-Order (HOLO) algorithm \cite{pa12a,chacon:2017} and the formulation is particularly suitable for multifrequency TRT problems. A well-informed reemission source is obtained from the discretely-consistent gray LO system. Because of the presence of the consistent, nonlinear LO system, the contribution from the collision physics (e.g., absorption-reemission) can be analytically integrated without imposing an explicit time-step constraint. The combination of the HOLO algorithm and the DP method makes the overall solution algorithm very efficient.

The rest of this paper is organized as follows. We first show the particle representation of the gray TRT equation in Section \ref{sec:method}. Then, our overall TRT solution strategy, based on the HOLO method, is presented. In Section \ref{sec:mg_extension}, we extend our DP algorithm to the multifrequency problem. Section \ref{sec:dp-holo-alg} shows the overall solution algorithm for the developed DP-HOLO method, and highlights the similarities and differences between IMC and MOC. In Section \ref{sec:numerical_results}, we demonstrate the efficacy of the DP algorithm, and we summarize our findings and future work in Section \ref{sec:conclusion}. 
%
%
\section{Methods/Algorithm}
\label{sec:method}
For simplicity, we begin with the 1D gray absorption-reemission TRT model,
\begin{align}
\frac{1}{c}\frac{\partial I}{\partial t} + \mu \frac{\partial I}{\partial x} + \sigma I &= \sigma
\frac{acT^4}{2},\label{eq:trt_gray1d}\\
\rho c_v \frac{\partial T}{\partial t} + \sigma acT^4-\sigma cE &= 0 .\label{eq:tmat_gray1d}
\end{align}
We will consider an extension to a frequency-dependent system in the next section. In particle-based methods (e.g., Monte Carlo) the distribution function (i.e., specific angular intensity, $I$) is represented as a collection of particles with weight $w_{p'}$ as follows,
\begin{equation}
I(x,\mu,t) = \sum_{p'}^{N_p} w_{p'}(t) \delta(x-x_{p'}(t))\delta(\mu-\mu_{p'}(t)), \label{eq:dp_intensity}
\end{equation}
where $x_{p'}(t),~\mu_{p'}(t)$ are the spatial position and direction of the particle $p'$ at time $t$, respectively. 
The evolution equation for the particle weight, $w_p$,  can be obtained by multiplying Eq.\ (\ref{eq:trt_gray1d}) by $\delta(x-x_{p}(t))\delta(\mu-\mu_{p}(t))$ and integrating over phase space to find:
\begin{equation}
\frac{d w_p}{d t} + \sigma c w_p = c Q(x_p, t),
\end{equation}
where, $Q(x_p, t)=\frac{\sigma ac T^4(x_p, t)}{2}$.  The formal solution for $w_p$ along the characteristics is,
\begin{equation}
w_p(t) = w_p(0)e^{-\int_0^t \sigma c dt'} + \int_0^t e^{-\int_{t'}^t \sigma c dt'' }cQ(x_p(t'),t')dt' .\label{eq:p_evolution}
\end{equation}
Note that the particle weight is always positive because the integrand in Eq.\ (\ref{eq:p_evolution}) is always non-negative. Due to the absence of an acceleration term in Eq.\ (\ref{eq:trt_gray1d}), the particle trajectory can be simply expressed as,
\begin{align}
  x_p(t) &= x_p(0) + \mu_p ct ,\label{eq:ch_x}\\
  \mu_p(t) &=\mu_p(0) .\label{eq:ch_mu}
\end{align}
The integral of Eq.\ (\ref{eq:p_evolution}) can be evaluated analytically  when the source is reconstructed with polynomials, as will be the case here. In the following subsections, we describe implementation and algorithmic details of the DP-HOLO method.
%
%
\subsection{HOLO solution strategy}
In order to perform the integral in  the particle-weight evolution equation, Eq.\ (\ref{eq:p_evolution}), the reemission source term must be known \textit{a priori}. For this, we employ a similar approach to the CMFD-accelerated MOC algorithm for nuclear reactor eigenvalue calculations, but, instead of using CMFD, we use a recently developed, High-Order, Low-Order (HOLO) algorithm \cite{pa12a, pa12b, chacon:2017}. The HOLO algorithm is an iterative, moment accelerated scheme, with the LO system defined by taking the first two moments of Eq.\ (\ref{eq:trt_gray1d}) together with the material temperature equation,
\begin{align}
\frac{\partial E}{\partial t} + \frac{\partial F}{\partial x} + \sigma cE = \sigma acT^4, \label{eq:0th_mom} \\
\frac{1}{c}\frac{\partial F}{\partial t} + \frac{c}{3} \frac{\partial E}{\partial x} + \sigma F = 0, \label{eq:1st_mom}\\
\rho c_v \frac{\partial T}{\partial t} + \sigma acT^4-\sigma cE = 0. \label{eq:lo_tmat}
\end{align}
Here, $F(x,t)=\int \mu I(x,\mu,t) d\mu$ is the (gray) radiative flux. Note that we have used the standard $P_1$ closure in Eq.\ (\ref{eq:1st_mom}) instead of the more consistent Eddington factor closure \cite{goldin:1967}. Solving Eqs.\ (\ref{eq:0th_mom}) and (\ref{eq:1st_mom}) introduces inconsistencies between HO and LO descriptions, not only at the discrete level, but also in the continuum.  Consistency between the HO and the LO system can be regained by adding a consistency term \cite{pa12a} that takes into account the mismatch in truncation errors and the transport effects. Note that there will be no consistency term in Eq.\ (\ref{eq:0th_mom}) because, as we will show, this equation is satisfied automatically in both HO and LO systems.
%

\subsubsection{Discrete representation of the LO system and HOLO consistency}
  In the following, we describe how the LO system  is discretized consistently with the HO system. We begin with Eq.\ (\ref{eq:0th_mom}), which is written discretely as,
  \begin{equation}
    \frac{E^{n+1,HO}_i-E^{n,HO}_i}{\Delta t_n} + \frac{\bar{F}^{n+1,HO}_{i+1/2}-\bar{F}^{n+1,HO}_{i-1/2}}{\Delta x_i} + \sigma_i c\bar{E}^{n+1,HO}_i = \sigma_i ac (T^{n+1}_i)^4 \label{eq:e-ho}
  \end{equation}
  where $T^{n+1}_i$ is the material temperature at cell $i$,  $\Delta t_n = t_{n+1}-t_n$, $\Delta x_i = x_{i+1/2}-x_{i-1/2}$, $\sigma_i = \sigma(T_i)$ and,
  \begin{align}
    E^{n+1,HO}_i &= \frac{1}{c\Delta x_i}\int_{x_{i-1/2}}^{x_{i+1/2}}\int_{-1}^{1} I(x,\mu,t_{n+1} )d\mu dx, \\
    \bar{E}^{n+1,HO}_i &= \frac{1}{c \Delta t_n\Delta x_i}\int_{t_n}^{t_{n+1}}\int_{x_{i-1/2}}^{x_{i+1/2}}\int_{-1}^{1} I(x,\mu,t )d\mu dx dt, \\
    \bar{F}^{n+1,HO}_{i\pm 1/2} &= \frac{1}{\Delta t_n}\int_{t_n}^{t_{n+1}} \int_{-1}^{1} \mu I(x_{i\pm 1/2},\mu,t) d\mu dt.
  \end{align}
  These quantities are tallied from particles during the HO step as described below. Discrete consistency between LO and HO system demands that they have the same local energy balance, Eq.\ (\ref{eq:e-ho}). Because the (time-discrete) LO system cannot solve for both $E^{n+1}$ and $\bar{E}^{n+1}$ simultaneously,  in order for the LO system to be consistent with the temporal integration of the HO system we rearrange Eq.\ (\ref{eq:e-ho}) as:
  \begin{equation}
    \frac{\bar{E}^{n+1,HO}_i-\bar{E}^{n,HO}_i}{\Delta t_n} + \frac{\bar{F}^{n+1,HO}_{i+1/2}-\bar{F}^{n+1,HO}_{i-1/2}}{\Delta x_i} + \sigma_i c\bar{E}^{n+1,HO}_i = \sigma_i ac (T^{n+1}_i)^4 + \mathcal{R}_i^{n+1}, \label{eq:e-r-ho}\\
  \end{equation}
    where $\mathcal{R}_i^{n+1}= \frac{\bar{E}^{n+1,HO}_i- \bar{E}^{n,HO}_i}{\Delta t_n} -\frac{E^{n+1,{HO}}_i-E^{n,HO}_i}{\Delta t_n} $ is the residual source term \cite{densmore:2015}.

    Next, we propose the following closure equations for partial radiative fluxes at faces $x_{i+1/2}$ [Eq.\ (\ref{eq:1st_mom})]:
  \begin{align}
    \frac{\bar{F}^{n+1,HO,+}_{i+1/2}-\bar{F}^{n,HO,+}_{i+ 1/2}}{c \Delta t_n} + \frac{c}{6} \frac{\bar{E}^{n+1,HO}_{i+1}-\bar{E}^{n+1,HO}_i}{\Delta x_{i+ 1/2}} + \sigma_{i+1/2}  \bar{F}^{n+1,HO,+}_{i+1/2} &= \gamma^{n+1,+}_{i + 1/2} c \bar{E}^{HO,n+1}_i, \label{eq:fp}\\
    \frac{\bar{F}^{n+1,HO,-}_{i+1/2}-\bar{F}^{n,HO,-}_{i+ 1/2}}{c \Delta t_n} - \frac{c}{6} \frac{\bar{E}^{n+1,HO}_{i+1}-\bar{E}^{n+1,HO}_i}{\Delta x_{i+ 1/2}} + \sigma_{i+1/2}  \bar{F}^{n+1,HO,-}_{i+1/2} &= \gamma^{n+1,-}_{i + 1/2} c \bar{E}^{HO,n+1}_{i+1},\label{eq:fm}
  \end{align}
  with
  \begin{align}
    \bar{F}^{n+1,HO,+}_{i+ 1/2} &= \frac{1}{\Delta t_n}\int_{t_n}^{t_{n+1}} \int_{0}^{1} \mu I(x_{i+ 1/2},\mu,t) d\mu dt,\label{eq:pradF_plus}\\
    \bar{F}^{n+1,HO,-}_{i+ 1/2} &= \frac{1}{\Delta t_n}\int_{t_n}^{t_{n+1}} \int_{-1}^{0} |\mu| I(x_{i+ 1/2},\mu,t) d\mu dt, \label{eq:pradF_minus}\\
    \bar{F}^{n+1,HO}_{i+ 1/2} &= \bar{F}^{n+1,HO,+}_{i+ 1/2} -\bar{F}^{n+1,HO,-}_{i+ 1/2} .
  \end{align}
 which are also evaluated from particles. Here, we have introduced the consistency term, $\gamma^{n+1,\pm}$, such that the moments of the HO solution always satisfy Eqs.\  (\ref{eq:fp}) and (\ref{eq:fm}).
  Subtracting Eq.\ (\ref{eq:fm}) from Eq.\ (\ref{eq:fp}) yields,
  \begin{align}
    \frac{\bar{F}^{n+1,HO}_{i+1/2}-\bar{F}^{n,HO}_{i+ 1/2}}{c \Delta t_n} &+ \frac{c}{3} \frac{\bar{E}^{n+1,HO}_{i+1}-\bar{E}^{n+1,HO}_i}{\Delta x_{i+ 1/2}} + \sigma_{i+1/2} \bar{F}^{n+1,HO}_{i+1/2}\nonumber \\
    &= \gamma^{n+1,+}_{i + 1/2} c \bar{E}^{HO,n+1}_i- \gamma^{n+1,-}_{i + 1/2} c \bar{E}^{HO,n+1}_{i+1} . \label{eq:ftot}
  \end{align}
Eq.\ (\ref{eq:ftot}) is the discrete LO closure equation. Finally, the discrete LO system for cell $i$ consists of the following set of equations.
  \begin{align}
    \frac{\bar{E}^{n+1,LO}_i-\bar{E}^{n,LO}_i}{\Delta t_n} &+ \frac{\bar{F}^{n+1,LO}_{i+1/2}-\bar{F}^{n+1,LO}_{i-1/2}}{\Delta x_i} + \sigma_i c\bar{E}^{n+1,LO}_i = \sigma_i ac (T^{n+1}_i)^4 + \mathcal{R}^{n+1}_i , \label{eq:e-lo}\\
    \frac{\bar{F}^{n+1,LO}_{i+1/2}-\bar{F}^{n,LO}_{i + 1/2}}{c \Delta t_n} &+ \frac{c}{3} \frac{\bar{E}^{n+1,LO}_{i+1}-\bar{E}^{n+1,LO}_i}{\Delta x_{i+ 1/2}} + \sigma_{i+1/2} \bar{F}^{n+1,LO}_{i+1/2} \nonumber \\
    &= \gamma^{n+1,+}_{i + 1/2} c \bar{E}^{n+1,LO}_i - \gamma^{n+1,+}_{i + 1/2} c \bar{E}^{n+1,LO}_{i+1}, \label{eq:ftot-lop}\\
    \frac{\bar{F}^{n+1,LO}_{i-1/2}-\bar{F}^{n,LO}_{i - 1/2}}{c \Delta t_n} &- \frac{c}{3} \frac{\bar{E}^{n+1,LO}_{i-1}-\bar{E}^{n+1,LO}_i}{\Delta x_{i- 1/2}} + \sigma_{i-1/2} \bar{F}^{n+1,LO}_{i-1/2} \nonumber \\
    &= - \gamma^{n+1,-}_{i - 1/2} c \bar{E}^{n+1,LO}_i + \gamma^{n+1,-}_{i - 1/2} c \bar{E}^{n+1,LO}_{i-1}, \label{eq:ftot-lom}\\
    \rho c_v \frac{T^{n+1}_i-T^{n}}{\Delta t_n} &+\sigma_i ac (T^{n+1}_i)^4- \sigma_i c \bar{E}^{n+1,LO}_i  =0 .\label{eq:t-lo}
  \end{align}

  Discrete consistency can be simply realized because the HO moments satisfy linear Eqs.\ (\ref{eq:e-r-ho}) and (\ref{eq:ftot}), given $T^{n+1}_i$ and $\gamma^{n+1,\pm}_{i+1/2}$. Thus, upon the convergence of $T^{n+1}$ and $\gamma^{n+1,\pm}_{i+1/2}$, and assuming consistency at the previous time-step, there is a unique solution which satisfies Eqs.\ (\ref{eq:e-r-ho}) and (\ref{eq:ftot}) simultaneously. Furthermore, it is also the only solution which satisfies  Eqns.\ (\ref{eq:e-lo})-(\ref{eq:t-lo}).

  \subsubsection{Energy conservation property}
After the nonlinear convergence between the HO and LO systems, summing Eqs.\  (\ref{eq:e-lo}) and (\ref{eq:t-lo}) gives the following local energy conservation,
  \begin{align}
&    \frac{\bar{E}^{n+1}_i- \bar{E}^{n}_i}{\Delta t_n} + \rho c_v \frac{T^{n+1}_i-T^{n}}{\Delta t_n}  + \frac{\bar{F}^{n+1}_{i+1/2}-\bar{F}^{n+1}_{i-1/2}}{\Delta x_i} -\mathcal{R}^{n+1}_{i}  \nonumber\\
    =&     \frac{E^{n+1,HO}_i-E^{n,HO}_i}{\Delta t} + \rho c_v \frac{T^{n+1}_i-T^{n}}{\Delta t_n}  + \frac{\bar{F}^{n+1}_{i+1/2}-\bar{F}^{n+1}_{i-1/2}}{\Delta x_i}\nonumber\\
    =&0. \nonumber
  \end{align}
Because we use a single value for the face radiative flux, $\bar{F}^{n+1}_{i\pm1/2}$, our HOLO scheme is conservative.
The overall DP-HOLO algorithm is summarized in Sec. \ref{sec:dp-holo-alg}.
%
%
\subsection{Tallying}
Eq.\ (\ref{eq:e-ho}) requires evaluation of three moments, $E^{n+1},~\bar{E}^{n+1},$ and $\bar{F}^{n+1}$, during a DP-HO step. Using the specific angular intensity definition of Eq.\ (\ref{eq:dp_intensity}), the end-of-time step radiation energy density for cell $i$, $E^{n+1}_i$, can be evaluated from particles as,
\begin{align}
  E^{n+1}_i &= \frac{1}{c \Delta x} \int_{x_{i-1/2}}^{x_{i+1/2}} \int_{-1}^{1} I(x,\mu, t_{n+1}) d\mu dx, \nonumber \\
   & = \frac{1}{c \Delta x}\sum_{p \in i} w_p (t_{n+1}) ,
\end{align}
which is merely the sum of the particle weights located in the cell $i$ at the end of the time step. We note that this interpolation strategy of the particle weight to the mesh corresponds to a ``top-hat'' function of width equal to the cell size, and is formally first order accurate \cite{birdsall:2005}.
The time-averaged radiation energy density at cell $i$ is defined as,
\begin{align}
  \bar{E}^{n+1}_i &= \frac{1}{c\Delta x_i \Delta t_{n}} \int_{x_{i-1/2}}^{x_{i+1/2}} \int_{-1}^{1} \int_{t_n}^{t_{n+1}} I(x,\mu,t) dt d\mu dx,\\
  &= \frac{1}{c\Delta x_i \Delta t_{n}} \int_{x_{i-1/2}}^{x_{i+1/2}} \int_{-1}^{1} \int_{t_n}^{t_{n+1}}\sum_p^{N_p} w_p (t) \delta(x-x_p(t))\delta(\mu-\mu_p(t)) dt d\mu dx,\\
  &=\frac{1}{c\Delta x_i \Delta t_{n}} \int_{x_{i-1/2}}^{x_{i+1/2}} \int_{t_n}^{t_{n+1}}\sum_p^{N_p} w_p (t) \delta(x-x_p(t)) dt dx .
  \end{align}
The above term is computed during the DP step as follows. First, within a time step, $t_n<t<t_{n+1}$, we partition the tallying step for each spatial cell, where the material properties are assumed to be constant. Without loss of generality, when a particle passes through the cell $i$ during interval $0<t<\Delta t$, the deposited particle weight in cell $i$ due to particle $p$ is given by,
\begin{align}
  \int_{0}^{\Delta t}\sigma_i c  w_p(t) dt &= \int_{0}^{\Delta t} \sigma_i c \left[ w_p(0) e^{-\sigma_i ct} + \int_{0}^{ t} e^{-\sigma_i c (t-t') }  cQ(x_p(t'),t') dt'\right]dt .\label{eq:dep}
\end{align}
Eq.\ (\ref{eq:dep}) is a more convenient tally choice because it directly represents absorbed energy. In addition, Eq.\ (\ref{eq:dep}) is required for evaluating energy-weighted opacities (see  Sec.\ \ref{sec:mg_extension}). We consider next the cases where the reemission source $Q(x,t)$ is either constant or linear in $x$, for which the source integral  in Eq.\ (\ref{eq:dep}) can be performed analytically. For a constant source, the particle weight at time $t$ can be expressed as,
\begin{align}
  w_p(t) &= w_p(0)e^{-\int_0^t \sigma_i c dt'} + \int_0^t e^{-\int_{t'}^t \sigma_i c dt'' }cQ(x_p(t'),t')dt'\nonumber\\
  &= w_p(0)e^{-\sigma_i ct} + \frac{Q_i}{\sigma_i}\left(1-e^{-\sigma_i ct}\right),
  \label{eq:w_const}
\end{align}
where $Q_i$ is the average reemission source in cell $i$. Note that for IMC with ``continuous energy deposition'' \cite{fl71}, the particle weight evolves with only the first term in Eq.\ (\ref{eq:w_const}). Consequently, the particle weight quickly vanishes in optically-thick cells, and surface radiative flux tallies become noisy \cite{densmore:2015}. On the other hand, it is easy to see that the particle weight for the DP method approaches $w_p \approx \frac{Q}{\sigma}= \frac{acT^4}{2}$ in optically-thick regions, which is the correct equilibrium limit and thus results in accurate surface radiative flux tallies.

Next, multiplying Eq.\ (\ref{eq:w_const}) by $\sigma_i c$ and integrating over the time interval $0<t<\Delta t$ yields 
an explicit expression for the deposited particle weight,
\begin{align}
  \int_{0}^{\Delta t}\sigma_i c  w_p(t) dt
  &= \int_{0}^{\Delta t}\sigma_i c \left[ w_p(0) e^{-\sigma_i ct} + \frac{Q_i}{\sigma_i}\left(1-e^{-\sigma_i ct} \right)\right]dt \nonumber \\
  &= w_p(0)\left(1-e^{-\sigma_i c \Delta t}\right)
  + Q_i c\Delta t - \frac{ Q_i }{\sigma_i} \left(1- e^{-\sigma_i c \Delta t} \right) \nonumber \\
  &= w_p(0)G(\tau)
  + Q_i c\Delta t - \frac{ Q_i }{\sigma_i} G(\tau) \label{eq:weight_dep_const},
\end{align}
where $\tau = \sigma_i c \Delta t$, and $G(\tau)= 1-e^{-\tau}$.
The first term in Eq.\ (\ref{eq:weight_dep_const}) corresponds to the deposition of the original particle weight. The second term corresponds to the weight gain due to the reemission source, and the third term corresponds to deposition of the reemission source  along the particle trajectory.   

For a (spatially) linear reemission source within a cell (the one employed here), $x_{i-1/2}<x<x_{i+1/2}$,  we may express the spatial variation of the reemission source as,
\begin{align}
  Q( x ) &= Q_i + (x-x_{i-1/2})\tilde{Q}_i .
\end{align}
Then, along the particle trajectory, $x_p(t) = x_0+\mu_p ct$, we can write the reemission source term as a linear polynomial in time,
\begin{align}
  Q(t) &= Q_i + (x(t)-x_{i-1/2})\tilde{Q}_i \nonumber\\
  &=Q_i -\tilde{Q}_i x_{i-1/2} + (x_0 + \mu_p ct) \tilde{Q}_i\nonumber\\
  &=Q_i +(x_0-x_{i-1/2})\tilde{Q}_i + \tilde{Q}_i\mu_p ct  \nonumber\\
  &= Q_0 + t\hat{Q} .
\end{align}
Accordingly, the particle weight at time $t$ can be written as,

\begin{align}
  w_p(t) &= w_p(0)e^{-\int_0^t \sigma_i c dt'} + \int_0^t e^{-\int_{t'}^t \sigma_i c dt'' }cQ(x_p(t'),t)dt'\nonumber\\
  &= w_p(0)e^{-\sigma_i ct} + \frac{Q_0}{\sigma_i}\left(1-e^{-\sigma_i ct}\right) + \frac{\hat{Q}t}{\sigma_i} - \frac{c \hat{Q}}{(\sigma_i c)^2}(1-e^{-\sigma_i ct}) ,
  \label{eq:w_linear}
\end{align}
and the weight deposition in the cell becomes,
\begin{align}
  \int_{0}^{\Delta t}\sigma_i c  w_p(t) dt
  &= \int_{0}^{\Delta t}\sigma_i c \left[ w_p(0)e^{-\sigma_i ct} + \frac{Q_0}{\sigma_i}\left(1-e^{-\sigma_i ct}\right) + \frac{\hat{Q}t}{\sigma_i} - \frac{c \hat{Q}}{(\sigma_i c)^2}(1-e^{-\sigma_i ct})\right]dt \nonumber \\
  &= w_p(0)G(\tau)
  + Q_0 c\Delta t - \frac{ Q_0 }{\sigma_i}G(\tau)+ \frac{c \hat{Q}\Delta t^2}{2} - \frac{\hat{Q}}{\sigma_i}\left( \Delta t - \frac{G(\tau)}{\sigma_i c}\right) . \label{eq:weight_dep_lin}
\end{align}
  This is the formula employed in this study, and is formally second-order accurate.

  Lastly, a surface-integrated, outgoing partial radiative flux [Eq.\ (\ref{eq:pradF_plus})] can be estimated as:
  \begin{align}
    \bar{F}^{n+1,+}_{i+1/2}A_{i+1/2} &= \frac{1}{\Delta t_n}\int_{t_n}^{t_{n+1}} \int_{A_{i+1/2}} \int_0^1 \mu I(x_{i+1/2},\mu,t) dA d\mu dt \nonumber \\
    & = \frac{1}{\Delta t_n} \int_{t_n}^{t_{n+1}} \int_{A_{i+1/2}} \int_0^1 \mu \sum_{p'}^{N_p} w_{p'}(t) \delta(x_{i+1/2}-x_{p'}(t))\delta(\mu-\mu_{p'}(t)) dA d\mu dt \nonumber \\
    & = \frac{1}{\Delta t_n} \int_{t_n}^{t_{n+1}} \int_0^1 \mu \sum_{p'}^{N_p} w_{p'}(t) \delta(x_{i+1/2}-x_{p'}(t))\delta(\mu-\mu_{p'}(t)) \left(\frac{A_{i+1/2}}{\mu}\right) d\mu dt \nonumber \\
    &= \frac{A_{i+1/2}}{\Delta t_n}  \sum_{p'}^{N_p} w_{p'}(t(x_p=x_{i+1/2}) )
  \end{align}
  where $A_{i+1/2}/\mu$ is the effective cross-sectional area.
%
%
\subsection{Boundary conditions}
There are two main types of boundary conditions in radiative transfer simulations: inflow boundary (including vacuum) and reflective boundary. In our DP algorithm, when a particle $p$ escapes from the boundary at time $t$, a new particle $p'$ enters the computational domain with the following attributes,
\begin{align}
  x_{p'}(t) &= x_{p}(t),\label{eq:bc_x}\\
  \mu_{p'}(t) &= -\mu_{p}(t),\label{eq:bc_mu}\\
w_{p'}(t) &=  
\begin{cases} 
w_{p}(t), \mathrm{~~reflective~ boundary}\\
\frac{ac T_{BC}^4}{2}, \mathrm{~~inflow~ boundary} .\label{eq:bc_w}
\end{cases}
\end{align}
where $T_{BC}$ is the inflow boundary temperature. As seen from Eq.\ (\ref{eq:bc_w}), the only difference between the two boundary conditions is the new particle weight. For either boundary conditions,  a new particle ``reflects'' back into the computational domain. In this way, the particle population remains constant unless particle re-sampling is performed.

  The boundary conditions for the LO system are provided in terms of the radiative fluxes. For example, the left boundary radiative flux at $x=0$ can be written as,
  \begin{align}
    \bar{F}^{n+1}_{1/2} = \bar{F}^{n+1,+}_{1/2}-\bar{F}^{n+1,-}_{1/2} .
  \end{align}
  Note that $\bar{F}^{n+1}_{1/2} \equiv \bar{F}^{n+1}(x=0)$.
  The partial radiative fluxes can be tallied while performing the HO step. Then,  we employ the following form of the LO flux boundary condition at the left boundary,
  \begin{align}
    \bar{F}^{n+1,LO}_{1/2} = \bar{F}^{n+1,HO,+}_{1/2}-\bar{F}^{n+1,HO,-}_{1/2}\frac{\bar{E}^{n+1,LO}_1}{\bar{E}^{n+1,HO}_1} . \label{eq:lo-bc0}
  \end{align}
  Similarly for the right flux boundary condition we use,
  \begin{align}
    \bar{F}^{n+1,LO}_{N+1/2} = \bar{F}^{n+1,HO,+}_{N+1/2} \frac{\bar{E}^{n+1,LO}_N}{\bar{E}^{n+1,HO}_N}-\bar{F}^{n+1,HO,-}_{N+1/2} . \label{eq:lo-bc1}
  \end{align}

\subsection{Particle initialization}

  Particle initialization requires specification of position, angle and weight. We define initial phase-space location of each particle  $(x_p,\mu_p)$ by uniformly dividing the number of initial particles located in cell $i$, with tensorial product. Number of space $N_{xi}$ and angular points $N_{\mu i}$ ($N_{pi} = N_{xi}N_{\mu i}$) within the cell are input parameters. Then, a particle initial phase-space location in the cell $i$ can be expressed as,
  \begin{align}
    x_p(t=0)   &= \frac{\Delta x_i}{N_{xi}}\left(j+\frac{1}{2}\right)+x_{i-1/2}, ~~j=1,N_{xi}, \label{eq:xp0}\\
    \mu_p(t=0) &= \frac{2}{N_{\mu i}}\left(m+\frac{1}{2}\right)-1, ~~m=1,N_{\mu i}. \label{eq:mup0}
  \end{align}
  In addition, each particle has an ``effective'' phase space volume of $dp = \frac{2 \Delta x_i}{N_{pi}}$.
To determine weight, in this work, we assume that the initial radiation distribution is isotropic in angle, and Planckian in frequency. Then, the initial radiation energy density can be described by the temperature, $T_0$ as,
  \begin{equation}
    E_0 = aT_0^4.
  \end{equation}
  Using Eq.\ (\ref{eq:dp_intensity})  and the definition of the cell-averaged radiation energy density yields,
  \begin{align}
    E_i(t=0) &=\frac{1}{c \Delta x_i} \int_{4\pi}\int_0^\infty \int_{x_{i-1/2}}^{x_{i+1/2}} I d\nu d\Omega dx \nonumber\\
    &= \frac{1}{c\Delta x_i}\sum_{p', \mathbf{r}_{p'}\in V_i}^{N_p} w_{p'}(t=0) . \label{eq:Enp1}
  \end{align}
Then, the initial particle weight in the cell $i$ becomes,
  \begin{equation}
    w_p(t=0) = \frac{c E_0\Delta x_i}{N_{pi}}, \label{eq:initial_weight}
  \end{equation}
  where $N_{pi}$ is the number of particles in cell $i$.  
%
\section{Multigroup Extension}
\label{sec:mg_extension}
Extending the DP algorithm for frequency-dependent problems is straightforward with a multifrequency approximation, in which the group-wise distribution function, $I_g$, is an integral value over the frequency range $\nu_{g-1}<\nu<\nu_{g}$,
\begin{equation}
  I_g(x,\mu,t) = \int_{\nu_{g-1}}^{\nu_g} I(x,\mu,\nu,t) d\nu .
  \label{eq:mf_def}
\end{equation}
With this definition, we can rewrite the TRT equation as,
\begin{align}
\frac{1}{c}\frac{\partial I_g}{\partial t} + \mu \frac{\partial I_g}{\partial x} + \sigma_g I &= \frac{\sigma_g
b_gacT^4}{2} = Q_g(x_p,t), ~g=1,...,N_g  ,\label{eq:trt_mf1d}
\end{align}
where $b_g$ is the normalized Planck function for group $g$. Following a similar procedure as in the previous section, we can write the particle weight evolution equation per frequency group as,
\begin{align}
w_p^g(t) = w_p^g(0)e^{-\int_0^t \sigma_g c dt'} + \int_0^t e^{-\int_{t'}^t \sigma_g c dt'' } cQ_g(x_p(t'),t')dt'. \label{eq:p_evolution_mf}
\end{align}
We emphasize here that there is a significant advantage of the DP method vs IMC for multifrequency problems in that the particle trajectory equations, Eqs.\ (\ref{eq:ch_x}) and (\ref{eq:ch_mu}), are independent of frequency and thus all frequency weights can be integrated simultaneously. Thus, we can provide phase space resolution with the same number of particles regardless of the number of frequency groups. Although each particle carries additional information, e.g., all group-wise particle weights, $w^g$, we can  reuse the ray-tracing evaluations for all frequency groups and thus significantly reduce computational effort. This will be demonstrated numerically in Sec.\ \ref{sec:numerical_results}.

  The gray LO system becomes,
  \begin{align}
    \frac{\bar{E}^{n+1,LO}_i-\bar{E}^{n,LO}_i}{\Delta t_n} &+ \frac{\bar{F}^{n+1,LO}_{i+1/2}-\bar{F}^{n+1,LO}_{i-1/2}}{\Delta x_i} + \sigma_{E,i} c\bar{E}^{n+1,LO}_i = \sigma_{p,i} ac (T^{n+1}_i)^4 + \mathcal{R}_i \label{eq:e-lo-mg}\\
    \frac{\bar{F}^{n+1,LO}_{i+1/2}-\bar{F}^{n,LO}_{i + 1/2}}{c \Delta t_n} &+ \frac{c}{3} \frac{\bar{E}^{n+1,LO}_{i+1}-\bar{E}^{n+1,LO}_i}{\Delta x_{i+ 1/2}} + \sigma_{R,i+1/2} \bar{F}^{n+1,LO}_{i+1/2} \nonumber\\
    &=  \gamma^{n+1,+}_{i + 1/2} c \bar{E}^{n+1,LO}_i - \gamma^{n+1,+}_{i + 1/2} c \bar{E}^{n+1,LO}_{i+1}\label{eq:ftot-lop-mg}\\
    \frac{\bar{F}^{n+1,LO}_{i-1/2}-\bar{F}^{n,LO}_{i - 1/2}}{c \Delta t_n} &- \frac{c}{3} \frac{\bar{E}^{n+1,LO}_{i-1}-\bar{E}^{n+1,LO}_i}{\Delta x_{i- 1/2}} + \sigma_{R,i-1/2} \bar{F}^{n+1,LO}_{i-1/2} \nonumber \\
    &= - \gamma^{n+1,\pm}_{i - 1/2} c \bar{E}^{n+1,LO}_i + \gamma^{n+1,\pm}_{i - 1/2} c \bar{E}^{n+1,LO}_{i-1}\label{eq:ftot-lom-mg}\\
    \rho c_v \frac{T^{n+1}_i-T^{n}}{\Delta t_n} &+\sigma_{P,i} ac (T^{n+1}_i)^4- \sigma_{E,i} c \bar{E}^{n+1,LO}_i  =0 \label{eq:t-lo-mg}
  \end{align}
  where the gray opacities are computed as:
  \begin{align}
    \sigma_{E,i} &= \frac{\sum_g{ \sigma_{g,i} E_{g,i}} }{\sum_g{E_{g,i}}}, \label{eq:sigmae} \\
    \sigma_{P,i} &= \frac{\sum_g \sigma_{g,i} B_g(T_i) }{\sum_g B_g(T_i)}, \label{eq:sigmap}\\
    \sigma_{R,i+1/2} &= \frac{\sum_g \frac{\partial B_g}{\partial T}|_{T_i}}{\sum_g\frac{1}{\sigma_{g,i}}\frac{\partial B_g}{\partial T}|_{T_i}} \label{eq:sigmar}
  \end{align}
  Here, we evaluate $\sigma_{g,i}$ at $T^{n}$, but we use the current radiation energy density and Planckian spectrum for the weights.
%
%
\section{Overall DP-HOLO Algorithm}
\label{sec:dp-holo-alg}
%
\begin{algorithm}[H]
  Input $N_{pi}, ~N_{\mu,i}$, and $T_0$ \;
  Initialize particle weight $w_p(0)$ from $T_0$ and Eq.\ (\ref{eq:initial_weight}) \;
  Initialize particle phase-space location $(x_p(0), \mu_p(0))$ with Eqs.\ (\ref{eq:xp0}) and (\ref{eq:mup0})\;
  Set $t_0=0$\;
  \For{$n=0,N$}
      {
        $t_{n+1}=t_n+\Delta t_n$\;
        Evaluate multigroup opacity $\sigma_g(T^{n}_{i})$ for $g=1,N_g$\;
        \textbf{\textit{LO predictor step}}\;
        Evaluate $\sigma_{E,i}$, $\sigma_{P,i}$, and $\sigma_{R,i\pm1/2}$ from Eqs.\ (\ref{eq:sigmae})-(\ref{eq:sigmar}) using $\bar{E}^{n,HO}_i$ and $T^{n}_i$\;
        Solve nonlinear Eqs.\  (\ref{eq:e-lo-mg})-(\ref{eq:t-lo-mg}) using boundary conditions Eqs.\ (\ref{eq:lo-bc0}) and (\ref{eq:lo-bc1}) to obtain predictor values $ T^{n+1}_{i,p} $\;
        
        \textbf{\textit{Source Reconstruction}}\;
        Reconstruct linear source term from $T^{n+1}_{i,p}$\;
        \textbf{\textit{HO solve}}\;
        Evolve weight of each particle using Eq.\ (\ref{eq:p_evolution_mf}) with $T^{n+1}_{i,p}$, and tally $\bar{E}^{n+1,HO}_{g,i}, \overline{\sigma_i E}^{n+1,HO}_{i},~ \bar{F}^{n+1,HO,\pm}_{i\pm 1/2}$\;
        
        \textbf{\textit{Closure Evaluation}}\;
        Evaluate $E^{n+1}_{i}$ from Eq.\ (\ref{eq:Enp1})\;
        Evaluate $\sigma_{E,i}^{n+1}$, $\sigma_{P,i}^{n+1}$, and $\sigma_{R,i\pm1/2}^{n+1}$ from Eqs.\ (\ref{eq:sigmae})-(\ref{eq:sigmar})\;
        Evaluate $\gamma^{n+1, \pm}_{i \pm 1/2}$ via Eq.\ (\ref{eq:fp}) and (\ref{eq:fm})\;
        Evaluate $\mathcal{R}_i= \frac{\bar{E}^{n+1,HO}_{i}- \bar{E}^{n,HO}_{i}}{\Delta t_n} -\frac{E^{n+1,HO}_{i}-E^{n,HO}_{i}}{\Delta t_n} $\;
        \textbf{\textit{LO corrector step}}\;
        Solve nonlinear {Eqs.\  (\ref{eq:e-lo-mg})-(\ref{eq:t-lo-mg})} using boundary conditions Eqs.\ (\ref{eq:lo-bc0}) and (\ref{eq:lo-bc1})
        with preconditioned Newton-Krylov method together with nonlinear elimination \cite{pa12b}\;
      }
\caption{Multigroup DP-HOLO predictor-corrector iterations} 
\label{al:holo}
\end{algorithm}

Algorithm \ref{al:holo} shows our multigroup DP-HOLO iterative scheme. In this work, we employ a predictor-corrector time-stepping \cite{pa14}. 
  In our current implementation of the DP-HOLO algorithm, the total number of particles remains constant during a simulation. Any particle that leaves the computational domain immediately reflects back into the domain with an appropriate weight. Optimization and particle-count adaptivity are left to the future work. During each time step, a particle carries the following attributes,
  \begin{enumerate}
  \item cell id ($i$),
  \item initial/current weight ($w_0/w$),
  \item initial/current location ($x_0/x$),
  \item initial/current direction ($\mu_0/\mu$),
  \item remaining time to the end of time step ($dt$),
  \item phase space volume ($dp$),
  \end{enumerate}
  We keep the initial particle state in order to allow HO solver restart during HOLO iterations within a time-step. Note that ``initial'' particle state refers to the state at the beginning of each time step, not the beginning of the simulation.

The LO system is solved iteratively via preconditioned Newton-Krylov method. We utilize the NOX solver package of Trilinos \cite{Trilinos-Overview}. Because the detail of the LO solver is presented in \cite{pa12b}, we only provide a brief summary in here. First, $k$-th Newton-step can be expressed as,
\begin{align}
  \mathcal{J}^k \delta \mathbf{U}^k &= -\mathcal{F}(\mathbf{U}^k).\\ \label{eq:newton}
  \mathbf{U}^{k+1} &= \mathbf{U}^k + \delta \mathbf{U}^k
\end{align}
Here, $\mathcal{F}(\mathbf{U}^k)$ is the nonlinear residual, evaluated by the current solution $\mathbf{U}^k$. $\mathcal{J}^k = \frac{\partial \mathcal{F}}{\partial \mathbf{U}}|_{\mathbf{U}^k}$ is the Jacobian matrix.
The analytical Jacobian matrix is formed at each Newton-step from Eqs. (\ref{eq:e-lo-mg})-(\ref{eq:t-lo-mg}), and solved via a multigrid preconditioned GMRES solver to the relative tolerance of $10^{-3}$. Nonlinear elimination \cite{lanzdron96, young03} of the material temperature and the radiative flux variables is employed in order to solve the LO system efficiently. Nonlinear elimination is relatively straight forward to implement in the TRT system due to absence of spatial coupling in the nonlinear term (absorption-reemission). The resulting Jacobian matrix for the radiation energy density resembles the Fleck-Cummings linearized equation \cite{fl71}. We use the nonlinear (relative) tolerance of $10^{-8}$ for checking convergence of the Newton iterations.
The computational cost for solving the LO system is essentially independent of the number of particles and time-step size, but depends on the number of the spatial cells. The LO solver typically requires less than 5 Newton iterations and less than 5 GMRES iterations per Newton iteration to converge. 

%
%
\subsection{Algorithm comparison with IMC and MOC}
In this subsection, we highlight algorithmic similarities and differences of the DP-HOLO from IMC and MOC.
First, all three are particle-based methods, and thus time-integration along particle trajectories can be performed analytically, unlike other grid-based methods such as $S_N$.

The first difference between IMC and DP-HOLO is the use of random numbers. As stated previously, our DP-HOLO algorithm does not employ random numbers. 
In DP-HOLO, each particle is deterministically generated at $t=0$, and it follows a straight trajectory until the end of simulation (or reflects back when the particle reaches the boundary). On the other hand, IMC particles are generated at each time step, based on the volumetric and boundary source terms. The phase-space location of the source particles is randomly sampled. 
In contrast, source contributions in DP-HOLO are accounted for in the particle weight evolution equation.

Second, IMC linearizes the absorption-reemission term, and resulting in the following linearized TRT equation \cite{fl71}:
\begin{align}
\frac{1}{c}\frac{\partial I}{\partial t} + \mu \frac{\partial I}{\partial x} + \sigma I = f ac T^4_n + \int_{-1}^{1} (1-f) \sigma I d\mu
\end{align}
where $f=\frac{1}{1+ \frac{4\sigma ac \Delta t T_n^3}{\rho c_v}}$ is the Fleck factor. IMC introduces an ``effective'' scattering term via linearization. The particle trajectory must be resampled when the particle undergoes scattering events (in addition to sampling of distance to the scattering/absorption events). 
It is known that this linearization introduces an additional stability constraint for the time-step size \cite{fl71,la87}. Instead,  DP-HOLO solves nonlinear absorption-reemission directly via the LO system, which allows it to follow the dynamical time scale of the problem (See Sec. \ref{sec:mpv}).  

Lastly, the evolution equation for the IMC particle weight between two events (e.g., absorption/scattering, boundary crossing) only takes into account exponential decay of the weight (the first term of RHS of Eq.\ (\ref{eq:p_evolution})). i.e.,
\begin{align}
w_p^{IMC}(t) = w_p^{IMC}(0)e^{-\int_0^t \sigma c dt'}. \label{eq:p_evolution_IMC}
\end{align}
Thus, IMC particles in optically thick (and cold) regions quickly lose their weight, possibly leading to noisy tallies of end-of-time-step radiation energy density, $E^{n+1}_i$.  

  DP-HOLO is closely related to the space-time-characteristics method (MOC) \cite{pandya:2009}, as both use the same space-time characteristics. However DP-HOLO is specifically designed to ensure discrete consistency and solve multigroup TRT problems. A (temporally and spatially) discretely consistent LO system is key to the successful application of the space-time characteristic method to TRT problems. 

%
%
\section{Numerical Examples}
\label{sec:numerical_results}
In order to demonstrate effectiveness of the DP-HOLO algorithm, we compare its accuracy and efficiency against the state-of-the-art Implicit Monte Carlo (IMC) method. We note that many accelerations and variance reduction algorithms, as well as novel code optimizations, exist for IMC that we have not considered in our comparisons. Nonetheless, given the several order-of-magnitude performance advantage of DP-HOLO vs. IMC, it is unlikely that such optimizations can alter the fundamental conclusion of this study. For all the results shown in this section, IMC employs a continuous energy deposition (implicit capture), a uniform particle sampling in each spatial cell, and frequency stratified sampling. {For DP-HOLO, all the results are obtained with the linear reemission source reconstruction, which has second-order convergence rate. Since we use a point particle (delta-function) to represent distribution functions (formally first-order accurate) and the linear reconstruction for the reemission source (formally second-order accurate), we expect the DP-HOLO algorithm will yield convergence between $O(N_p)$ and $O(N_p^2)$, depending on the dominant error components.}

  In the subsequent section, we present convergence plots of IMC and DP-HOLO. Errors are measured by comparing against their own reference solutions obtained from very large number of particles (1280 and 8192 particles per cell for IMC and DP-HOLO, respectively) while fixing the spatial and temporal discretizations. Thus, we are only measuring algorithmic convergence with respect to the number of particles used. We expect the computed error:
  \begin{equation}
    error ~= ~\frac{\sum_i |T_{ref,i}-T_i|}{T_{BC}} \label{eq:error}
  \end{equation}
to exhibit a proper measure of convergence rate of both IMC and DP-HOLO. Furthermore, convergence plots are presented with actual CPU time rather than the number of particles because the CPU cost per particle differs significantly between IMC and DP-HOLO, while the CPU time is almost directly proportional to the number of particles in both.  Finally, we use 8 particles to discretize the angular variable for all DP-HOLO simulations.
%
%
\subsection{Gray Marshak Wave}
The first example we consider is a gray Marshak wave problem. In this example, we use two different sets of material parameters that correspond to optically thin and thick regions. Table \ref{tab:marshak-prop} shows the problem definitions. In both examples, the system is initially in equilibrium, $T_m = T_r=T_0$, where $T_m$ and $T_r=(\frac{E}{a})^{1/4}$ are the material and radiation temperatures, respectively. An incoming radiation source is applied at $x=0$. The opacity is a function of the material temperature as follows,
\begin{equation}
\sigma(T) = \frac{\rho \alpha}{T^3}\label{eq:opacity-gray-marshak} .
\end{equation}
\begin{table}[!h]
\centering
  \begin{tabular}{|c|c|c|}
	\hline
             &  Problem 1 & Problem 2\\
             &  (Thick) & (Thin)\\
	\hline 
        $x$ [cm]        & 0.25        & 2.0 \\
        $t_f $ [s]    & $2\times 10^{-8}$ & $5\times 10^{-8}$\\
        $\rho$ [g/$\mathrm{cm}^3$]  & 1.0        & 1.0\\ 
        $c_v$   [erg/g-eV] & $3\times 10^{12}$ & $1.3874 \times 10^{11}$ \\
        $T_{bc}$ [eV]& 1000        &150 \\
        $T_0$    [eV]& 0.025       & 0.025\\
        $\alpha$ [$\mathrm{eV}^3-\mathrm{cm}^2$/g]& $1.0 \times 10^{12}$ & $1.0\times 10^6$ \\
        $\Delta x $ [cm] & 0.005 &0.025 \\
	\hline
\end{tabular}
\caption{Problem specifications for the Marshak wave problems}
\label{tab:marshak-prop}
\end{table}

\begin{figure}[!htb]
  \centering
  \subfigure[IMC solution, Tm]
  {
      \includegraphics[clip,width=3.5in]{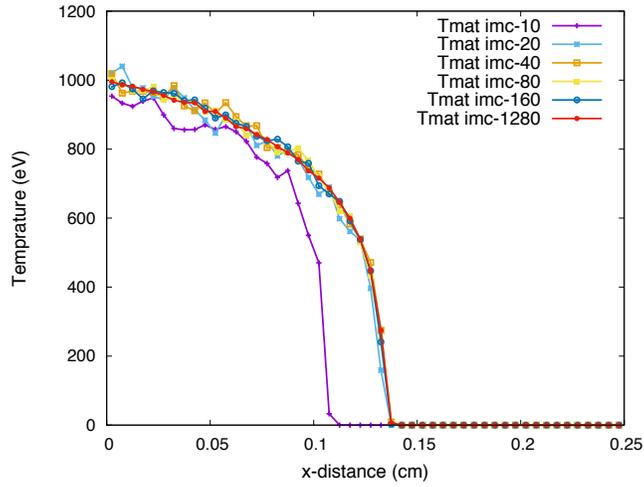}
      \label{imc-gray-marshak1-Tm}
  }
  \subfigure[DP solution, Tm]
  {
      \includegraphics[clip,width=3.5in]{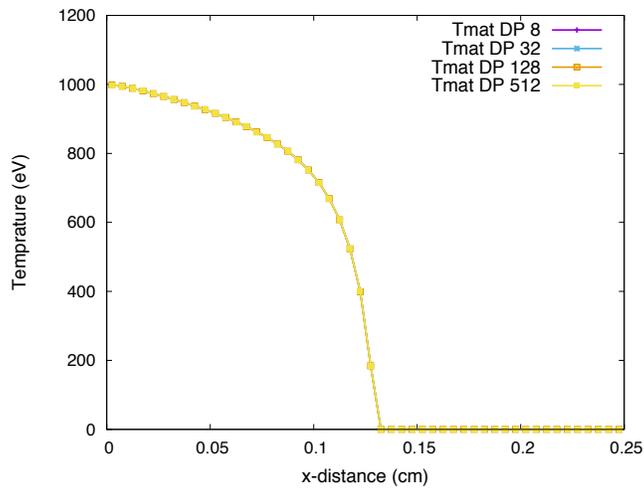}
      \label{dp-gray-marshak1-Tm}
  }
  \caption{Material temperature profiles for gray optically thick Marshak wave problem at  $t=2\times 10^{-8}$s.}
  \label{fig:marshak1-Tm}
\end{figure}

\begin{figure}[!htb]
  \centering
  \subfigure[IMC solution, Tr]
  {
      \includegraphics[clip,width=3.5in]{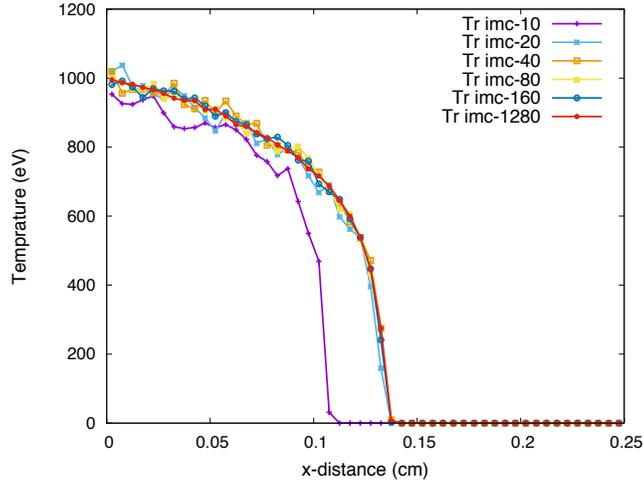}
      \label{imc-gray-marshak1-Tr}
  }
  \subfigure[DP solution, Tr]
  {
      \includegraphics[clip,width=3.5in]{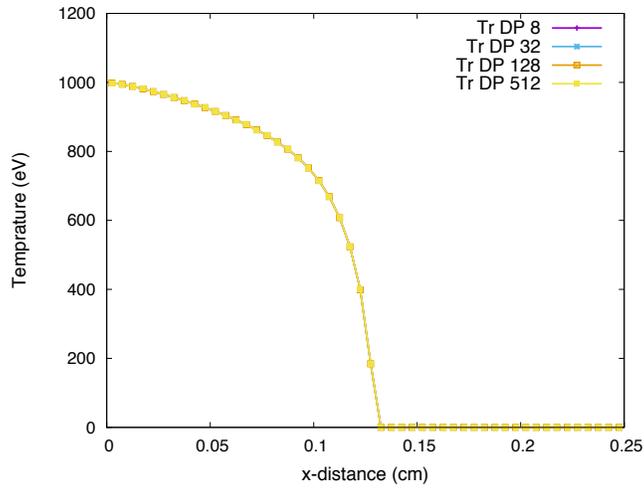}
      \label{dp-gray-marshak1-Tr}
  }
  \caption{ Radiation temperature profiles for gray optically thick Marshak wave problem  $t=2\times 10^{-8}$s.}
  \label{fig:marshak1-Tr}
\end{figure}

\begin{figure}[!htb]
  \centering
  \subfigure[IMC solution, Tm]
  {
      \includegraphics[clip,width=3.5in]{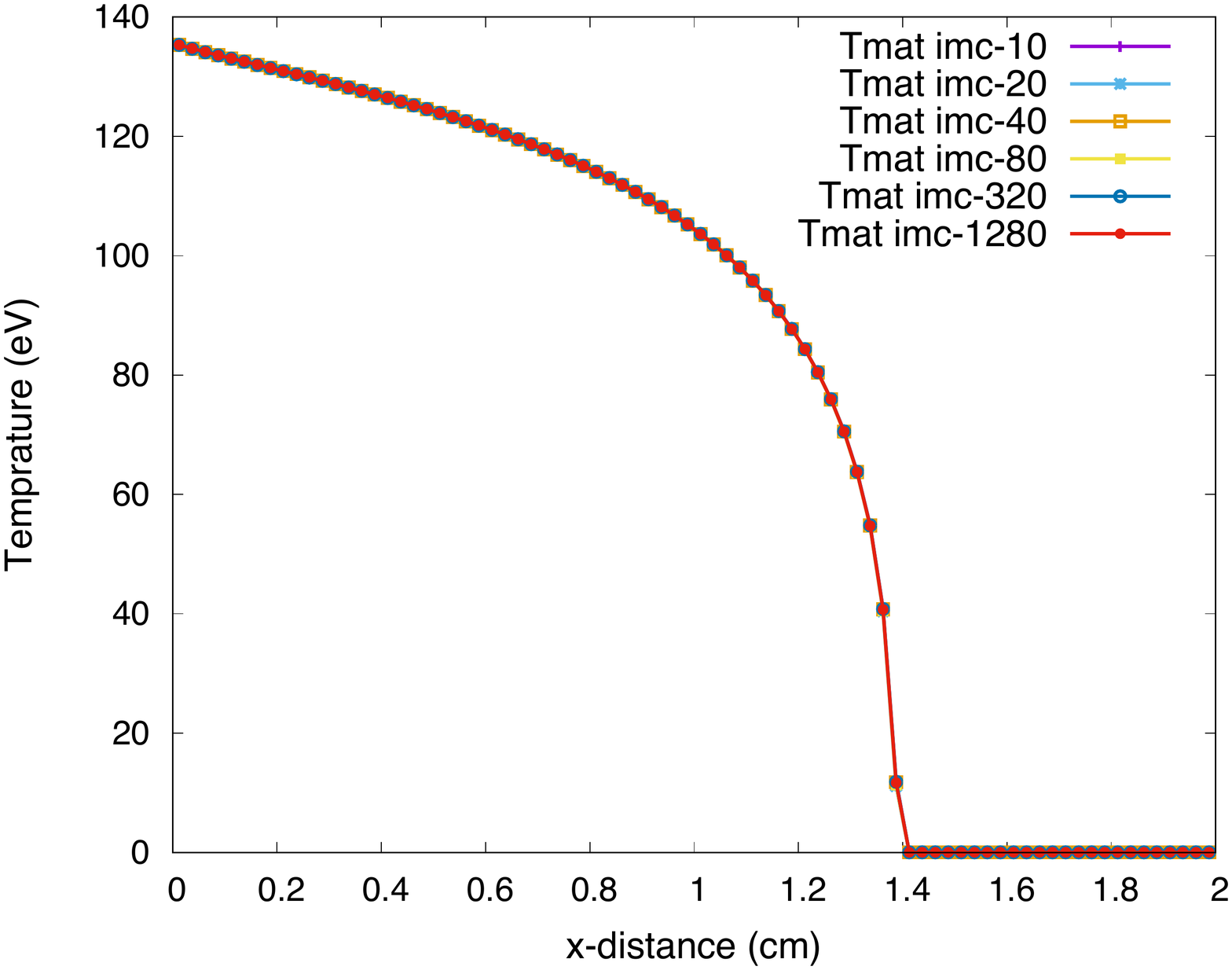}
      \label{imc-gray-marshak2-Tm}
  }
  \subfigure[DP solution, Tm]
  {
      \includegraphics[clip,width=3.5in]{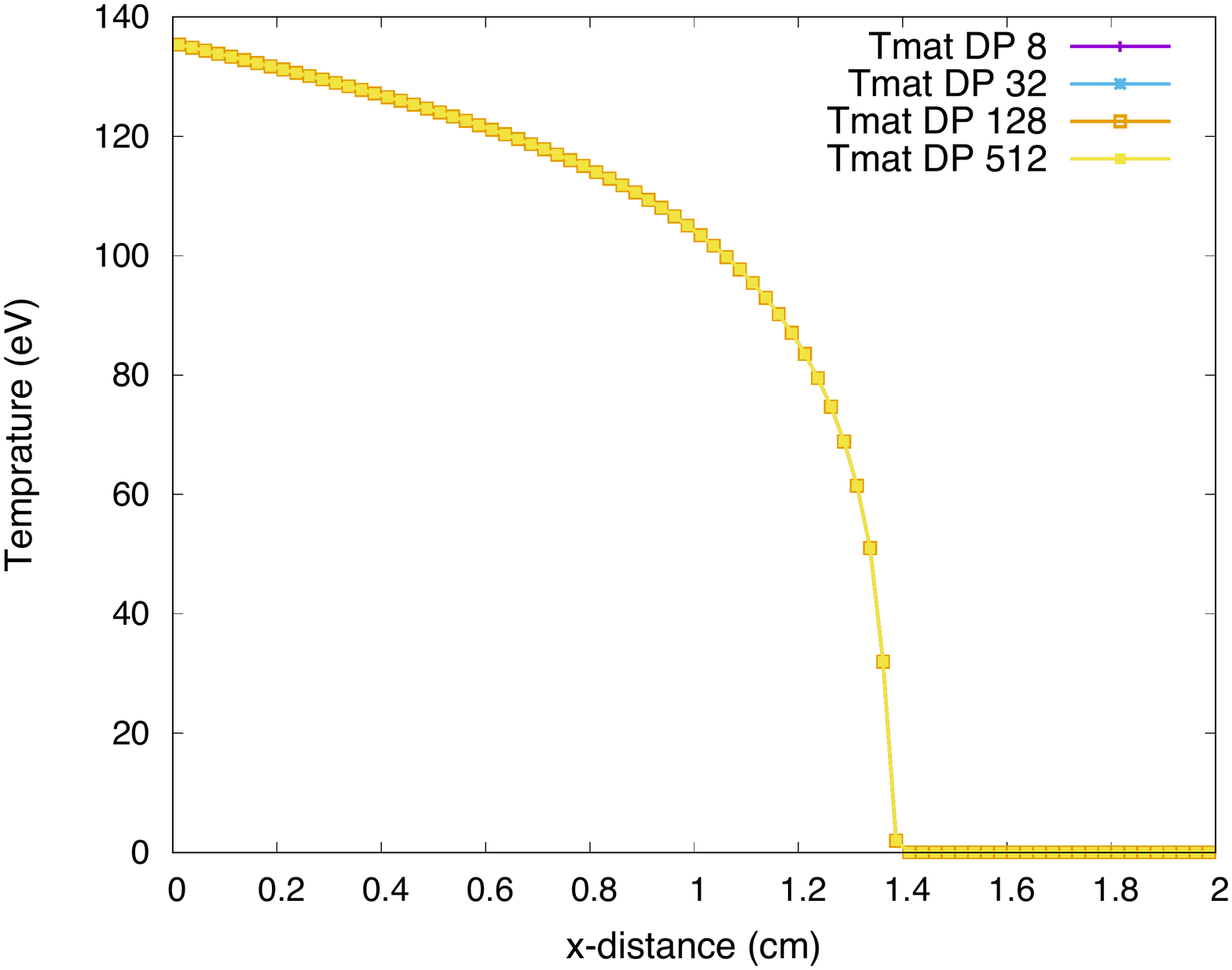}
      \label{dp-gray-marshak2-Tm}
  }
  \caption{Material Temperature profiles of gray optically thin Marshak problem at $t=5\times 10^{-8}$s.}
  \label{fig:marshak2-Tm}
\end{figure}

\begin{figure}[!htb]
  \centering
  \subfigure[IMC solution, Tr]
  {
      \includegraphics[clip,width=3.5in]{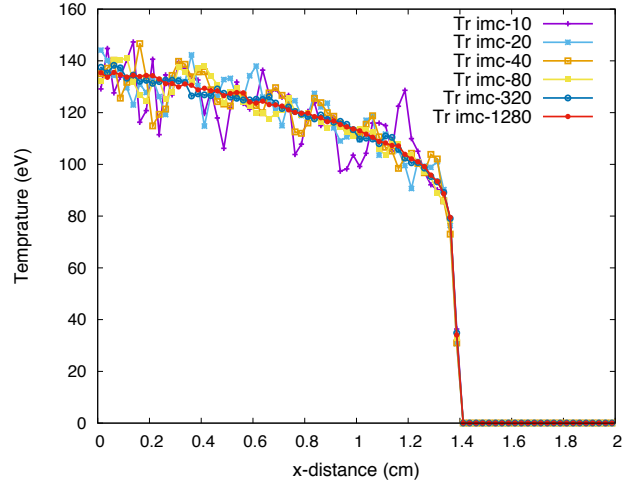}
      \label{imc-gray-marshak2-Tr}
  }
  \subfigure[DP solution, Tr]
  {
      \includegraphics[clip,width=3.5in]{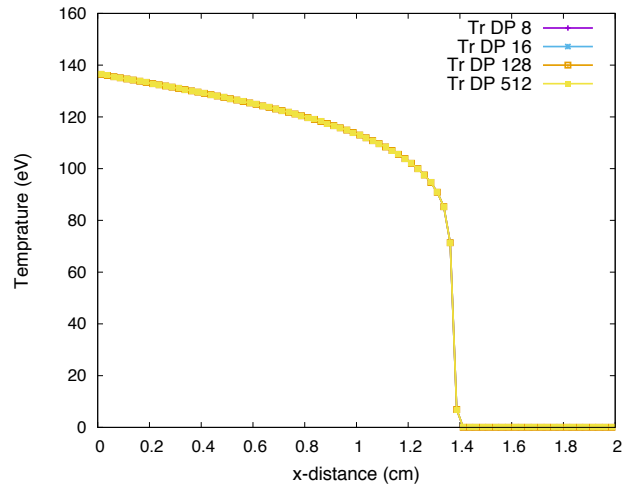}
      \label{dp-gray-marshak2-Tr}
  }
  \caption{Radiation temperature profiles of gray optically thin Marshak problem at $t=5\times 10^{-8}$s.}
  \label{fig:marshak2-Tr}
\end{figure}

Figs.\ \ref{fig:marshak1-Tm}-\ref{fig:marshak2-Tr} show the material and radiation temperature profiles computed with various particle resolutions. The value shown in the legends indicates the sampled number of particles  per cell (i.e., the number of reemission source particles per cell in IMC, number of initial particles per cell in DP). Initial time step sizes are $10^{-12}$ and $10^{-11}$ s for Problem 1 and Problem 2, respectively. Time step sizes are increased by a factor of $1\%$ and $5\%$ until they reach the maximum time step size of $10^{-10}$s.

\begin{figure}[!htb]
  \centering
  \includegraphics[height=3in]{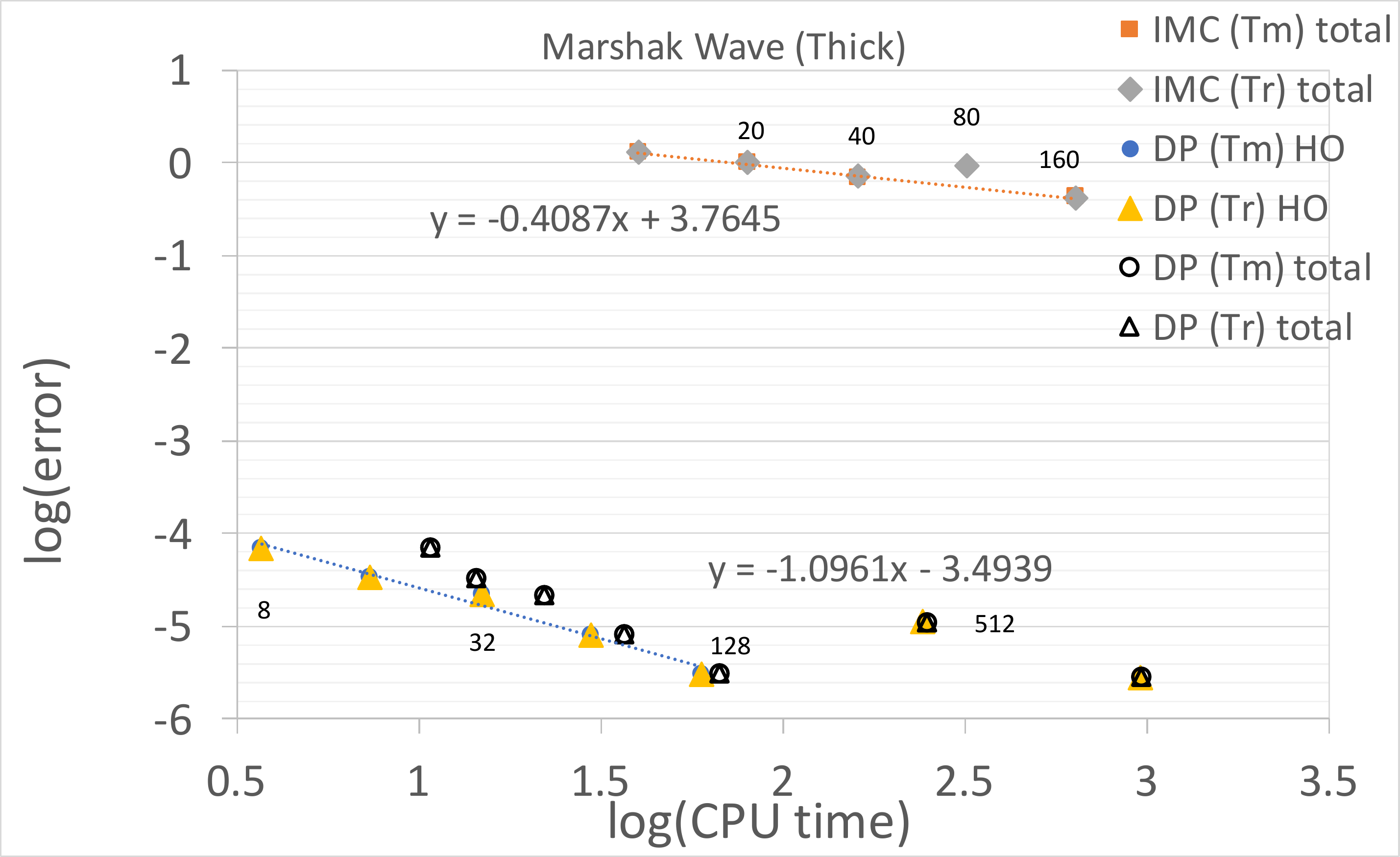}
  \caption{Error vs CPU time for the gray optically thick Marshak wave problem. Values by the data points indicate the number of particles per cell used in simulation in Figs.\ \ref{fig:marshak1-Tm} and \ref{fig:marshak1-Tr}. The HO CPU time of DP-HOLO and the total CPU time of IMC are proportional to the number of particles. The regression fit for DP-HOLO is evaluated using the HO CPU time.}
\label{fig:marshak1-error}
\end{figure}

The host medium for problem 1 is optically thick, and the radiation and material temperatures stay in equilibrium throughout the simulation. For the optically thick case, material and radiation temperature are essentially identical. The IMC solutions [Figs.\ \ref{imc-gray-marshak1-Tm} and \ref{imc-gray-marshak1-Tr}] exhibit a noticeable statistical noise in all cases but the one with 1280 particles per cell. On the other hand, the DP-HOLO solutions [Figs.\ \ref{dp-gray-marshak1-Tm} and \ref{dp-gray-marshak1-Tr}] show no significant noise regardless of the number of particles used. Fig.\ \ref{fig:marshak1-error} is an efficacy plot for this problem. We use the term ``efficacy'' to indicate required CPU time for a certain accuracy.   As can be seen from Fig.\ \ref{fig:marshak1-error}, errors in material and radiation temperatures are identical, as expected. We have plotted two sets of data for DP-HOLO. The first one is error vs.\ HO CPU time (blue and yellow data points), and the other is error vs.\ total CPU time (black data points). Note that in the total CPU time for IMC and the HO solver CPU time for DP-HOLO are almost proportional to the number of particles used.  Thus, it can be seen that the IMC simulations exhibit a typical $O(\sqrt{N_p})$ convergence rate, while the convergence rate of DP-HOLO appears to be  $O(N_p)$ (until the error is dominated by other sources such as the temporal and spatial discretizations). For DP-HOLO, the computational cost of the LO solver is about 7 sec, which corresponds to $0.7\%$ to $ 65.6\% $ of the total CPU time depending on the number of particles as shown in Table \ref{tab:cpu-marshak1}.

{
\begin{table}[!h]
\centering
  \begin{tabular}{|c|c|c|}
	\hline
         \# of particles    &  LO CPU time [s]  & HO CPU time [s]\\
	\hline 
        8    & 7.1 (65.6) & 3.7 (34.4)\\
        16   & 7.1 (48.9) & 7.4 (51.1)\\
        32   & 7.2 (32.6) & 14.9 (67.4)\\
        64   & 7.2 (19.4) & 29.7 (80.6)\\
        128  & 7.2 (10.8) & 59.5 (89.2)\\
        512  & 7.3 (2.9)  & 241.3 (97.1)\\
        2048 & 7.4 (0.7)  & 858.3 (99.3)\\
	\hline
\end{tabular}
\caption{CPU time distribution of DP-HOLO in gray optically thick Marshak wave problem. The values in parenthesis indicates \% CPU time with respect to the total CPU time.   }
\label{tab:cpu-marshak1}
\end{table}
}

\begin{figure}[!htb]
  \centering
  \includegraphics[height=3in]{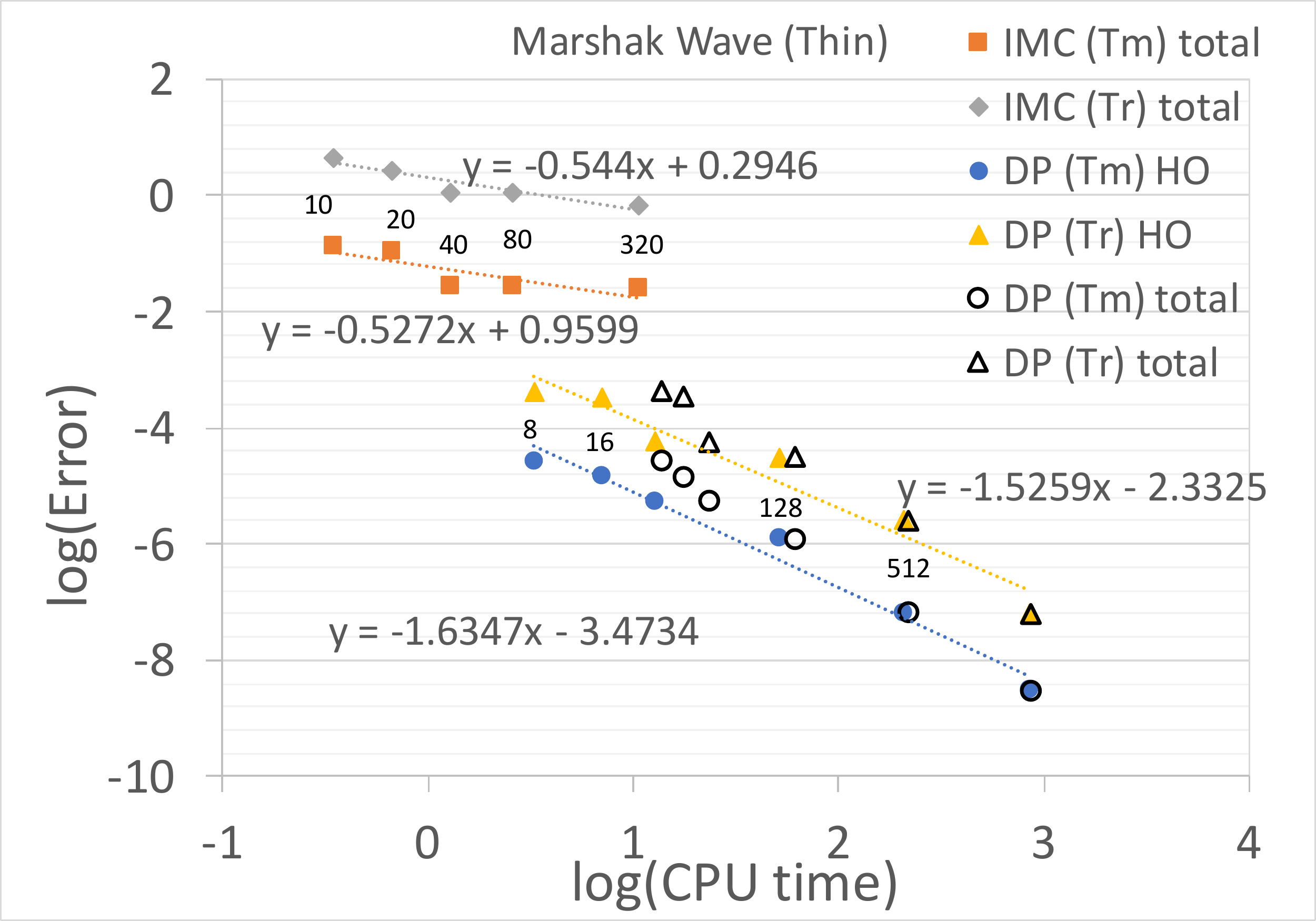}
  \caption{Error vs a CPU time  for the gray optically thin Marshak wave problem. Values by the data points indicate the number of particles per cell used in simulation in Figs.\ \ref{fig:marshak2-Tm} and \ref{fig:marshak2-Tr}. The HO CPU time of DP-HOLO and the total CPU time of IMC are proportional to the number of particles. The regression fit for DP-HOLO is evaluated using the HO CPU time.}
\label{fig:marshak2-error}
\end{figure}

Problem 2 has smaller opacity and incoming radiation temperature, and the material and radiation temperatures are decoupled and differ by about $40$ eV at the wave front.  The material temperature profiles do not exhibit statistical noise (unlike problem 1)  as seen from Fig.\ \ref{imc-gray-marshak2-Tm}. However, the IMC radiation temperature profile still shows significant statistical noise, seen in Fig.\ \ref{imc-gray-marshak2-Tr}, while DP radiation temperature solutions are free from the noise for any number of particles (Fig.\ \ref{dp-gray-marshak2-Tr}). Fig.\ \ref{fig:marshak2-error} shows the efficacy plot for this problem.  Similar to the previous case (Fig.\ \ref{fig:marshak1-error}), the IMC method converges close to $O(\sqrt{N_p})$, while the DP-HOLO method converges between $O(N_p)$ and  $O(N_p^2)$. Data in this figure is arranged in the same way as in Fig.\ \ref{fig:marshak2-error}.  The computational cost of the LO solver is  about 10.5 sec, which corresponds to 1.3 \% to 76.4\% of the total CPU time as shown in Table \ref{tab:cpu-marshak2}.  For both optically thin and thick cases, the DP-HOLO method delivers efficacies many orders of magnitude larger than IMC.

{
\begin{table}[!h]
\centering
  \begin{tabular}{|c|c|c|}
	\hline
         \# of particles    &  LO CPU time [s]  & HO CPU time [s]\\
	\hline 
        8    & 10.5 (76.4) & 3.3 (23.6)\\
        16   & 10.5 (59.7) & 7.1 (40.3)\\
        32   & 10.7 (45.3) & 12.9 (54.7)\\
        128  & 10.5 (17.0) & 51.4 (83.0)\\
        512  & 10.7 (4.9)  & 206.5 (85.1)\\
        2048 & 10.8 (1.3)  & 850.2 (98.7)\\
	\hline
\end{tabular}
\caption{CPU time distribution of DP-HOLO in gray optically thin Marshak wave problem. The values in parenthesis indicates \% CPU time with respect to the total CPU time.   }
\label{tab:cpu-marshak2}
\end{table}
}

%

%

\clearpage
\subsubsection{Stability with large time-step}
\label{sec:mpv}
It is well known that the Fleck-Cummings linearization used in IMC can produce nonphysical material temperature solutions when the time step becomes sufficiently large \cite{fl71,la87}. This unphysical phenomenon is often referred to as the ``violation of the maximum principle.'' When the maximum principle is violated, the material temperature becomes higher than the boundary (incoming) temperature, and stagnates the radiation wave. The only mitigation in the IMC method is to use a smaller time-step size. On the other hand, because of the nonlinearly consistent LO system, our DP-HOLO algorithm does not violate the maximum principle.

\begin{figure}[!htb]
  \centering
  \subfigure[IMC solution]
  {
      \includegraphics[clip,width=3.5in]{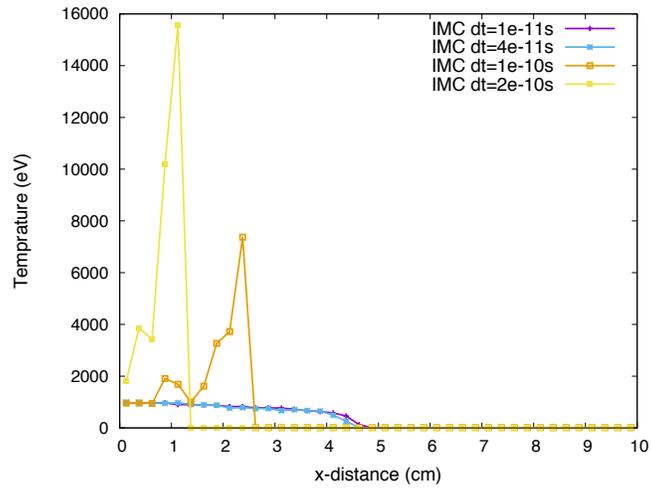}
      \label{imc-gray-marshak-Tm-mpv}
  }
  \subfigure[DP solution]
  {
      \includegraphics[clip,width=3.5in]{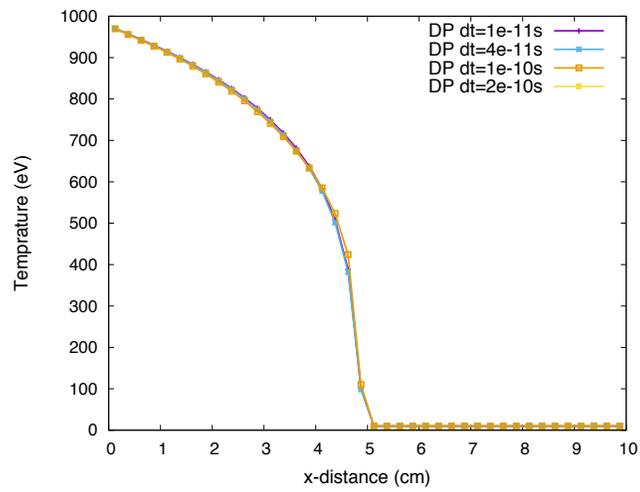}
      \label{dp-gray-marshak-Tm-mpv}
  }
  \caption{Material Temperature profiles of gray Marshak problem at $t=1\times 10^{-9}$s.}
  \label{fig:marshak-mpv}
\end{figure}

In order to demonstrate this effect, we run the gray Marshak wave problem with time steps of various sizes.
\begin{table}[!h]
\centering
  \begin{tabular}{|c|c|c|}
	\hline 
$x$ [cm]        & 10.0 \\
$t_f $ [s]    & $1\times 10^{-9}$\\
$\rho$ [g/$\mathrm{cm}^3$]  & 1.0\\
$c_v$   [erg/g-eV] & $1\times 10^{10}$\\
$T_{bc}$ [eV]& 1000 \\
$T_0$    [eV]& 10.0\\
$\alpha$ [$\mathrm{eV}^3-\mathrm{cm}^2$/g]& $1.0 \times 10^{9}$\\
	\hline
\end{tabular}
\caption{Problem specifications for the Marshak wave problems}
\label{tab:marshak-mpv-prop}
\end{table}
Fig.\ \ref{fig:marshak-mpv} shows the material temperature profile with the different time step sizes. It is clear that the IMC solution with large time step (e.g., $\Delta t > 10^{-10}$s) exhibits unphysical temperature spikes and a slow-moving wave front. On the other hand, the DP solution is virtually indistinguishable for all time step sizes.

%
\clearpage
%
%
\subsection{Larsen MG}

The next example we consider is a multifrequency Marshak wave problem (a.k.a Larsen's problem \cite{larsen:88}). This problem consists of three regions; region 1: $0<x<2.0$ cm, region 2: $2.0<x<3.0$ cm, and region 3: $3.0<x<4.0$ cm. We use spatial mesh size of $\Delta x=0.05$ cm. Each region has the following frequency-dependent opacity,

\begin{equation}
  \sigma(\nu,T) = \rho \alpha \frac{1-e^{-h\nu/kT}}{(h\nu)^3},
\end{equation}
where $\alpha$ is set to $10^9$, $10^{12}$, and $10^9$ for the regions 1, 2, and 3, respectively. The density and heat capacity are set to $1.0$ [g/cc] and $5.109 \times 10^{11}$ [erg/g-eV]. The system is initially in thermal equilibrium at $1.0$ eV and we apply a $10^3$ eV Planckian surface source at x = 0.0 cm to start the transient. Each simulation was run to $t = 6 \times 10^{−10}$ s with $\Delta t=2 \times 10^{-12}$s. Frequency groups are split uniformly in the log-scale between $10^{-2}$ eV and $10^5$ eV.

\begin{figure}[!htb]
  \centering
  \subfigure[IMC solution]
  {
      \includegraphics[clip,width=3.in]{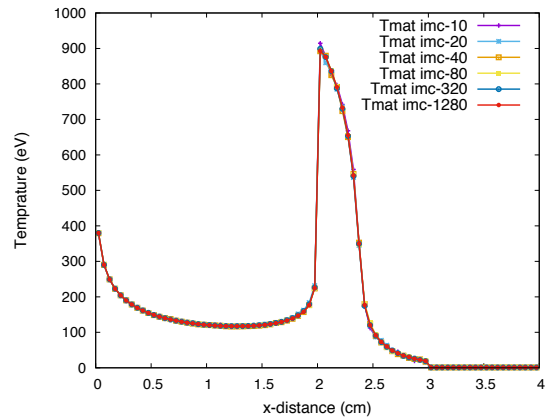}
      \label{larsen-imc-Tm}
  }
  \subfigure[DP solution]
  {
      \includegraphics[clip,width=3.in]{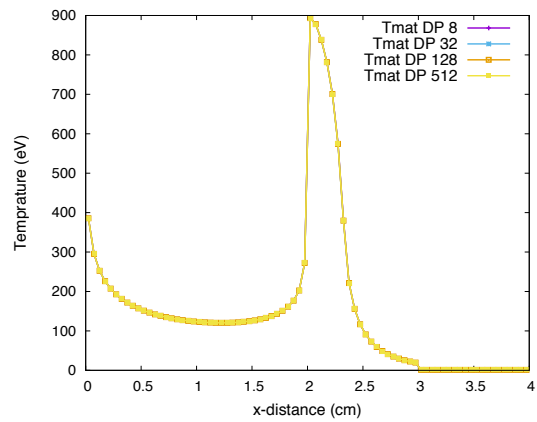}
      \label{larsen-dp-Tm}
  }
  \caption{Material temperature profiles for Multifrequency Larsen's problem.}
  \label{fig:larsen-Tm}
\end{figure}

\begin{figure}[!htb]
  \centering
  \subfigure[IMC solution]
  {
      \includegraphics[clip,width=3.in]{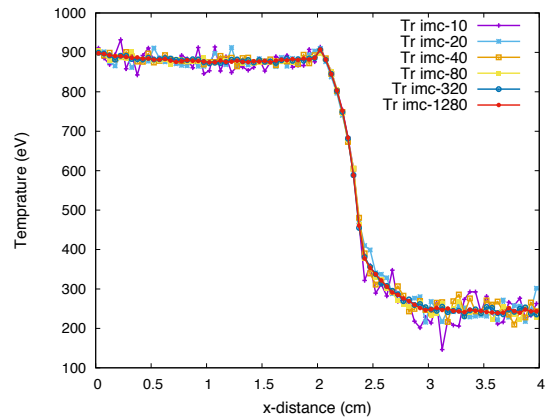}
      \label{larsen-imc-Tr}
  }
  \subfigure[DP solution]
  {
      \includegraphics[clip,width=3.in]{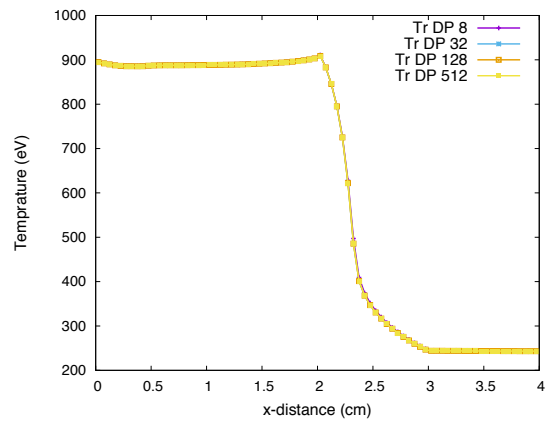}
      \label{larsen-dp-Tr}
  }
  \caption{Radiation temperature profiles for Multifrequency Larsen's problem.}
  \label{fig:larsen-Tr}
\end{figure}

\begin{figure}[!htb]
  \centering
  \includegraphics[height=3in]{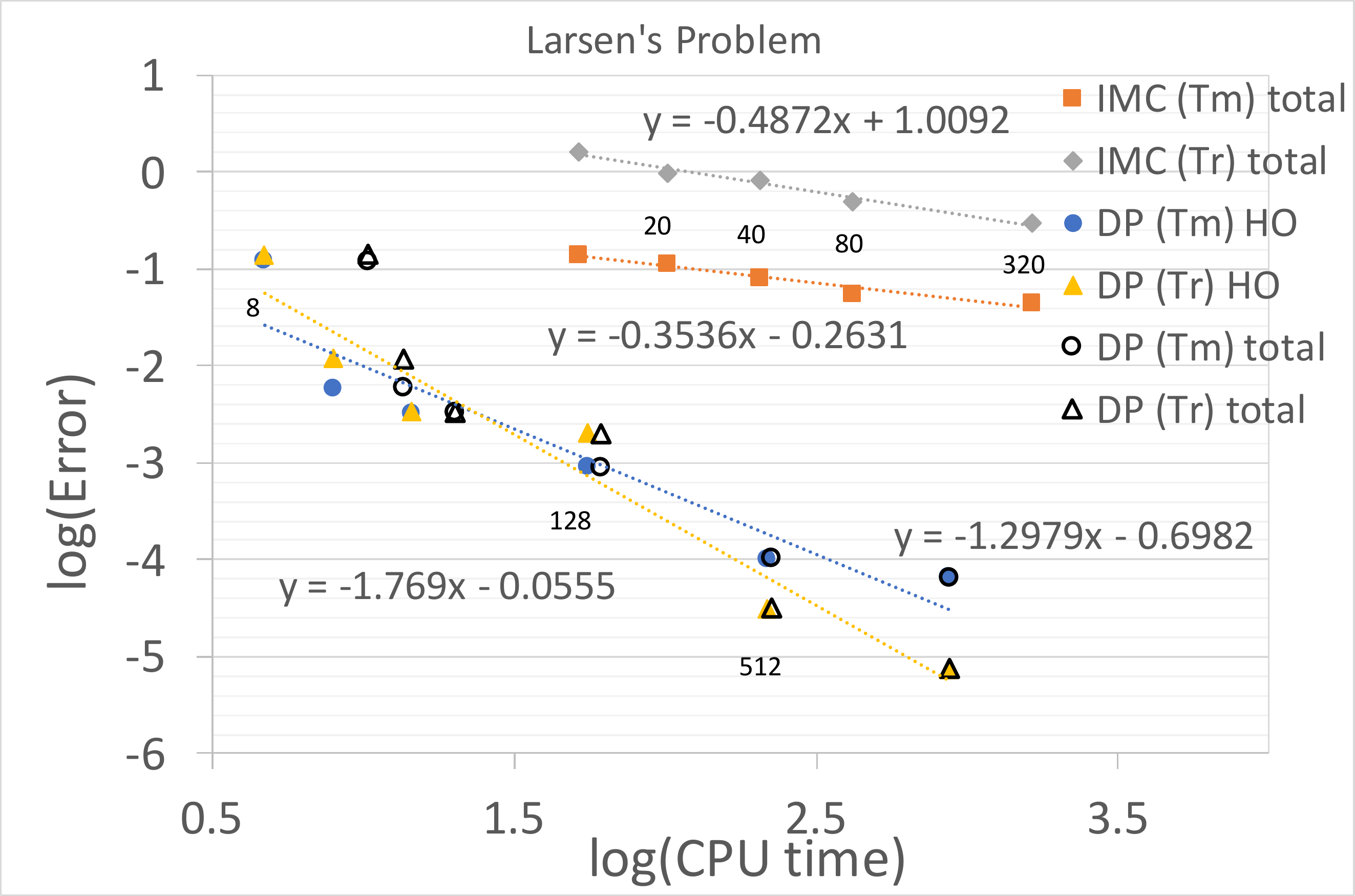}
  \caption{Error vs CPU time for the Larsen's problem. Values by the data points indicate the number of particles per cell used in simulation in Figs.\ \ref{fig:larsen-Tm} and \ref{fig:larsen-Tr}. The HO CPU time of DP-HOLO and the total CPU time of IMC are proportional to the number of particles. The regression fit for DP-HOLO is evaluated using the HO CPU time.}
\label{fig:larsen-error}
\end{figure}

Figs.\ \ref{fig:larsen-Tm} and \ref{fig:larsen-Tr} show the material and radiation temperature profiles computed with various numbers of particles. Compared to the material temperature profile, the radiation temperature is quite noisy when IMC is used. No noisy solutions  are observed with the DP-HOLO results. Fig.\ \ref{fig:larsen-error} shows the efficacy plot. Similarly to the previous examples, the convergence rate for the radiation temperature is approximately $O(\sqrt{N_p})$ and between $O(N_p)$ and $O(N_p^2)$ for IMC and DP-HOLO, respectively. Finally, the computational cost of the LO solver is about 6 sec, which corresponds to 0.7 $\%$ to 55 $\%$ of the total CPU time as shown in Table \ref{tab:cpu-larsen}.

{
\begin{table}[!h]
\centering
  \begin{tabular}{|c|c|c|}
	\hline
         \# of particles    &  LO CPU time [s]  & HO CPU time [s]\\
	\hline 
        8    & 5.7 (54.6) & 4.7 (45.4)\\
        16   & 5.7 (41.5) & 8.0 (58.5)\\
        32   & 5.8 (28.6) & 14.4 (71.4)\\
        128  & 6.0 (10.0) & 55.4 (90.0)\\
        512  & 6.0 (2.7)  & 217.8 (97.3)\\
        2048 & 6.1 (0.7)  & 867.8 (99.3)\\
	\hline
\end{tabular}
\caption{CPU time distribution of DP-HOLO in multifrequency Larsen's problem. The values in parenthesis indicates \% CPU time with respect to the total CPU time.   }
\label{tab:cpu-larsen}
\end{table}
}

\clearpage
%
%
\subsection{Multigroup Efficacy}
Lastly, we present the efficacy of the DP-HOLO algorithm in multifrequency problems.
First we simply check the algorithmic scaling with respect to the number of groups with a problem with frequency-independent opacity.
Here we use the material property of the optically thin gray Marshak wave problem presented previously (e.g., Eq.\ (\ref{eq:opacity-gray-marshak}), $\alpha=10^6$).

\begin{figure}[htb]
  \centering
  \includegraphics[height=3in]{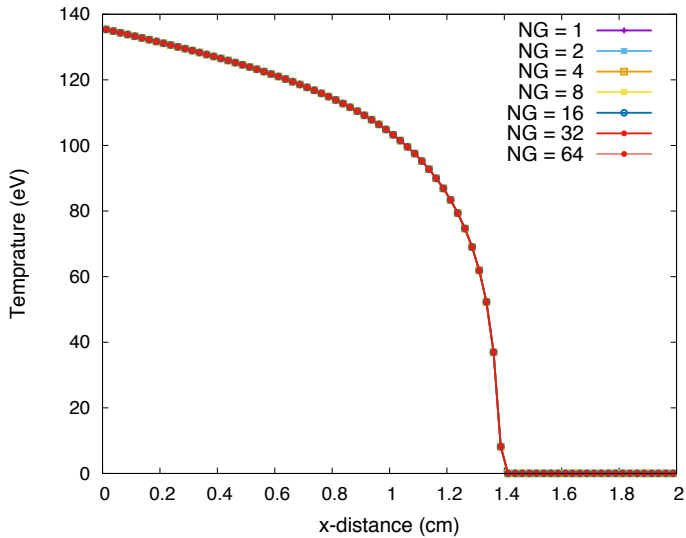}
  \caption{Material temperature profile of the Marshak wave problem with different numbers of groups.}
\label{fig:mf-marshak-ng}
\end{figure}

Fig.\ \ref{fig:mf-marshak-ng} shows the material temperature profile of the problem with different numbers of groups, which clearly indicates the solution (and hence the physics) is independent of the group structure. Table \ref{tab:marshak-g-relative-cpu} shows  relative CPU increase as a function of the number of groups. The relative CPU time for the IMC simulations in this problem scales linearly with the number of groups. On the other hand, it can be seen that the total CPU time of the DP-HOLO calculation with 64 groups only increases by a factor of 5.2 compared to the single-frequency group case. Furthermore, the CPU cost of the LO (gray) system is almost independent of the number of groups, and the HO CPU time increases only about a factor of 19 for 64 groups. This increase is significantly smaller than the group number increase of 64.

\begin{table}[!h]
  \centering
    \begin{tabular}{|c|c|c|c|c|}
      \hline
      & \multicolumn{4}{c|}{$CPU(NG)/CPU(NG=1)$} \\
      \hline
      NG & IMC & DP total & DP HO & DP LO \\
      \hline
      4  & 3.4  & 1.3 & 1.8  & 1.1\\
      8  & 6.9  & 1.5 & 2.9  & 1.1\\
      16 & 13.9 & 2.1 & 5.2  & 1.1\\
      32 & 27.5 & 3.1 & 9.7  & 1.1\\
      64 & 54.6 & 5.2 & 18.5 & 1.1\\
      \hline
    \end{tabular}
    \caption{{Relative CPU increase for frequency-independent Marshak problem.}}
    \label{tab:marshak-g-relative-cpu}
\end{table}

\begin{figure}[htb]
  \centering
\includegraphics[clip, height=3in]{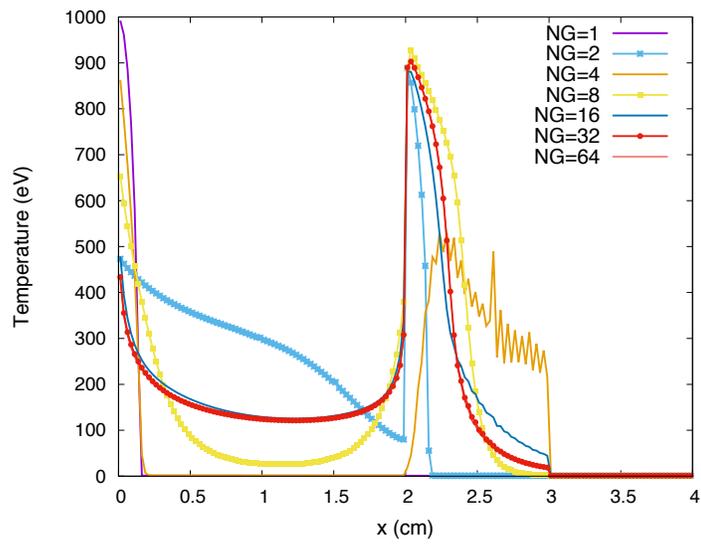}
\caption{Material temperature profile with different number of groups}
\label{fig:larsen-dp-g-Tmat}
\end{figure}

Next we use the Larsen's problem configuration. Fig.\ \ref{fig:larsen-dp-g-Tmat} shows material temperature profiles for different numbers of groups at $t=6 \times 10^{-10}s$. It shows that, due to the frequency-dependent opacity, the material temperature profiles change significantly when the frequency variable is not well-resolved. (The solution has converged for 64 groups.) Similar to Table \ref{tab:marshak-g-relative-cpu}, Table \ref{tab:larsen-g-relative-cpu} shows the relative CPU time increase as a function of groups.
In this problem, IMC does not scale linearly with the number of groups. This is mostly due to the fact that a single group simulation of this problem becomes optically very thick, and the IMC particles do not travel far (very small cost per particle), which results in artificially small CPU time for the single-group case. On the other hand, the scaling of the DP-HOLO simulations is almost identical to the previous case. This indicates that the CPU cost of DP-HOLO algorithm is almost independent of the physics regimes. 

  \begin{table}[!h]
    \centering
    \begin{tabular}{|c|c|c|c|c|}
      \hline
         & \multicolumn{4}{c|}{$CPU(NG)/CPU(NG=1)$} \\
      \hline
      NG & IMC & DP total & DP HO & DP LO \\
      \hline
      4  & 61.2  & 1.2 & 1.9  & 1.0\\
      8  & 58.4  & 1.8 & 3.5  & 1.3\\
      16 & 89.1  & 2.1 & 5.4  & 1.2\\
      32 & 167.5 & 3.2 & 10.0 & 1.4\\
      64 & 338.4 & 5.4 & 19.0 & 1.7\\
      \hline
    \end{tabular}
    \caption{Relative CPU increase in Larsen's problem.}
    \label{tab:larsen-g-relative-cpu}
  \end{table}

%
\clearpage
\section{Conclusion}
\label{sec:conclusion}
We have proposed a moment-accelerated deterministic particle algorithm for multifrequency thermal radiative transfer problems. The DP-HOLO method is closely related to the method of characteristics employed in nuclear reactor calculations. Unlike Implicit  Monte Carlo, the particle weight evolution takes into account the reemission source. An explicit expression of reemission source is obtained from a discretely consistent, moment-based low-order description. With numerical examples, we have demonstrated that the DP-HOLO method results in accurate material and radiation temperatures with a very small number of particles compared to IMC. 
Furthermore, the DP-HOLO method converges between $O(N_p)$ and  $O(N_p^2)$, as opposed to $O(\sqrt{N_p})$ for IMC. As a result, DP-HOLO has efficacies (accuracy per unit of cost) several orders of magnitude larger than IMC.  Due to the ability of the DP-HOLO particles to evolve weights for all frequencies simultaneously, the advantage of the DP method over IMC increases with multifrequency problems. Furthermore, the relative CPU time increase of DP-HOLO simulation is independent of the physics regime. Ongoing work includes an extension of the algorithm to  two-dimensional problems, as well as curvilinear geometries. Future work will focus on developing a new discretization scheme for the LO system that preserves the asymptotic diffusion limit. 

\section{Acknowledgment}
This work was performed with support of the Laboratory Directed Research and Development program at Los Alamos National Laboratory under US government contract DE-AC52-06NA25396 for Los Alamos National Laboratory, which is operated by Los Alamos National Security, LLC, for the US Department of Energy.
\section{Reference}
\bibliographystyle{elsarticle-harv}
\bibliography{bibs/refs}

\end{document}